\pdfoutput=1

\documentclass[%
 reprint,
 twocolumn,
 amsmath,amssymb,
 aps,
 floatfix
]{revtex4-1}

\usepackage{dcolumn}
\usepackage{bm}
\usepackage{hyperref}
\usepackage{natbib}
\usepackage[pdftex]{graphicx}
\usepackage{amsfonts}
\usepackage{mathrsfs}
\usepackage[english]{babel}
\usepackage{epsfig}
\usepackage{epstopdf}
\usepackage{babelbib}
\usepackage{color}
\usepackage{tcolorbox}
\usepackage{tabularx}
\usepackage{array}
\usepackage{colortbl}
\tcbuselibrary{skins}
\usepackage{comment}
\usepackage{array}
\usepackage{mathtools}

\newcommand{\tj}[6]{ \begin{pmatrix}
       #1 & #2 & #3 \\
       #4 & #5 & #6 
\end{pmatrix}}

\newcommand*{\textoverline}[1]{$\overline{\hbox{#1}}\m@th$}

\begin{document}

\title{From discrete to continuous description of spherical surface charge distributions}

\author{An\v ze Lo\v sdorfer Bo\v zi\v c}
\email{anze.bozic@ijs.si}
\affiliation{Department of Theoretical Physics, Jo\v zef Stefan Institute, Jamova 39, SI-1000 Ljubljana, Slovenia}

\date{\today}

\begin{abstract}
The importance of electrostatic interactions in soft matter and biological systems can often be traced to non-uniform charge effects, which are commonly described using a multipole expansion of the corresponding charge distribution. The standard approach when extracting the charge distribution of a given system is to treat the constituent charges as points. This can, however, lead to an overestimation of multipole moments of high order, such as dipole, quadrupole, and higher moments. Focusing on distributions of charges located on a spherical surface -- characteristic of numerous biological macromolecules, such as globular proteins and viral capsids, as well as of inverse patchy colloids -- we develop a novel way of representing spherical surface charge distributions based on the von Mises-Fisher distribution. This approach takes into account the finite spatial extension of individual charges, and leads to a simple yet powerful way of describing surface charge distributions and their multipole expansions. In this manner, we analyze charge distributions and the derived multipole moments of a number of different spherical configurations of identical charges with various degrees of symmetry. We show how the number of charges, their size, and the geometry of their configuration influence the behavior and relative importance of multipole magnitudes of different order. Importantly, we clearly demonstrate how neglecting the effect of charge size leads to an overestimation of high-order multipoles. The results of our work can be applied to construct analytical models of electrostatic interactions and multipole expansion of charged particles in diverse soft matter and biological systems.
\end{abstract}

%\keywords{Suggested keywords}

\maketitle

\section{\label{sec:intro}Introduction}

It is hard to underestimate the importance of charge and the resulting electrostatic interactions in various soft matter and biological systems. These include protein-protein and protein-polyelectrolyte interactions, viral capsid assembly and stability, interactions and crystallization of inverse patchy colloids, and drug delivery and cellular uptake of nanoparticles~\cite{Holm2012,Bianchi2017,Bianchi2017b,Siber2012,Bai2016}. Electrostatic interactions are also highly tunable and consequently enable a controllable and tunable assembly of charged particles~\cite{Bianchi2017,Bianchi2014}. The control over electrostatic effects can be achieved either by varying particle size and the size of their patches of charge, or by changing the properties of the surrounding electrolyte -- most importantly, its salt concentration and $pH$ value~\cite{Bianchi2014,Barisik2014,Kusters2015,Sabapathy2017,ALB2017a,Krishnan2017,Nap2014,Abrikosov2017}. What is more, the charge on biological macromolecules can be in principle also regulated via induced mutations, changing the nature and charge of their amino acid composition~\cite{Ni2012}.

Experimental observations and numerical simulations of electrostatic effects in these systems are often supplemented by analytical models~\cite{Holm2012,Bianchi2017,Bianchi2017b,Siber2012,Warshel2006}. In a first approximation, the total charge on a particle -- be it a colloid or a macromolecule -- can account for a large amount of its electrostatic behavior. But while such treatment of particles as homogeneously charged is customary, non-uniform charge effects often play a significant role that cannot be neglected~\cite{Adar2017,Grant2001}. For instance, both charge heterogeneity (patchiness) and charge fluctuation reduce the electrostatic repulsion between proteins or protein aggregates, eventually even giving way to attraction~\cite{Grant2001,ALB2013a,Li2017,Li2015,Vega2016}. Likewise, heterogeneity of charge is a determining factor in the aggregation and crystallization of inverse patchy colloids, as well as in their interaction with polyelectrolytes~\cite{Bianchi2017b,Blanco2016,Bianchi2014,Dempster2016,Yigit2015,Yigit2017}. Recent experiments have also revealed a long-range attraction between overall neutral surfaces, locally charged in a mosaic-like structure of positively and negatively charged patches~\cite{Silbert2012,Perkin2006,Meyer2005}.

Due to the typical size of colloidal and molecular systems, and the sheer number of atoms and charges involved in them, effective coarse-grained representations of their interaction potentials are vital for the modeling of such systems~\cite{Hoppe2013}. A common way of describing charge heterogeneities in particles and reducing their complexity is the multipole expansion of particles' surface charge distributions. This approach presents an efficient way of describing surface charges as continuous patches, easing the description of the rich set of surface charge patterns embedded in proteins and charged patchy particles~\cite{Blanco2016,Stipsitz2015}. In addition to determining and classifying the electrostatic multipole moments of different proteins~\cite{ALB2017a,Felder2007,Nakamura1985}, multipole expansion has also been widely used to explore protein-protein, protein-ligand, and colloidal interactions~\cite{Abrikosov2017,Blanco2016,Hoppe2013,Paulini2005,Parimal2014}, predict the electrophoretic mobility of proteins~\cite{Kim2006}, and to provide a representation of both the protein structure~\cite{Gramada2006} and of the symmetry of viral capsids~\cite{Lorman2007,Lorman2008}.

In obtaining a multipole representation of a particle's surface charge distribution, the charges on the particle are typically treated as point charges represented by Dirac $\delta$ functions. Similarly, the patches of charge on inverse patchy colloids are often considered to cover an exact surface area of the colloid, defined by sharp edges. While common, both descriptions are known to have multipole expansions where it is difficult to achieve an accurate representation of a surface charge distribution with a finite number of multipole terms, due to the Dirac $\delta$ and Heaviside step functions involved~\cite{ALB2013a,ALB2011}. And while an arbitrary cutoff can in principle be chosen, e.g., by representing a surface charge distribution only by its monopole, dipole, and quadrupole moments, this leaves open the question of accuracy and relevance of high-order multipole moments.

Here, we present a novel way of constructing spherical surface charge distributions based on the von Mises-Fisher distribution, taking into account the finite extent of individual charges on a given particle. We derive the expression for the multipole moments of thusly constructed distributions, yielding a simple yet elegant form which can be used to study how the number and size of charges as well as the geometry of their configuration on a particle influences the relative relevance of multipole moments of different order. The derived model presents an improvement in the description of the multipole representation of any number of charges on a spherical particle, with a simplicity which nonetheless allows it to serve as a more accurate input for analytical models of electrostatic effects in systems of globular proteins, viral capsids, and charged patchy colloids.

\section{Constructing spherical surface charge distributions}

We consider a point charge $q_k e_0$, located on a unit sphere of radius $R=1$ at a position $\mathbf{r}_k=(R,\vartheta_k,\varphi_k)=(R,\Omega_k)$, written in spherical coordinates; $e_0$ is the elementary charge. The contribution of the point charge to the total surface charge distribution on the sphere, when written in terms of the Dirac $\delta$ function, is
\begin{equation}
\label{eq:delta}
\sigma_\delta(\Omega)=\frac{q_k e_0}{R^2}\times\delta(\Omega-\Omega_k),
\end{equation}
normalized so that $\int\sigma_\delta(\Omega)\,\mathrm{d}V=q_k e_0$. Such a description, while standard, can cause difficulties when describing a contribution of many charges to the surface charge distribution and then expanding it in terms of multipoles. Specifically, the multipole coefficients of the distribution converge poorly, as an infinite sum over spherical harmonics is required to accurately represent the Dirac $\delta$ function.

In order to remedy this, we now represent a point charge $q_k e_0$ with a normal distribution on a sphere instead, writing its contribution to the total surface charge distribution as
\begin{equation}
\label{eq:vmf}
\sigma_\mathrm{vMF}(\Omega)=\frac{q_k e_0}{R^2}\times f(\Omega\,|\,\Omega_k,\lambda_k),
\end{equation}
where the function $f(\Omega\,|\,\Omega_k,\lambda)$ is the von Mises-Fisher (vMF) distribution on a unit sphere in three dimensions~\cite{Mardia2009},
\begin{equation}
f(\Omega\,|\,\Omega_k,\lambda)=\frac{\lambda}{4\pi\sinh\lambda}\,\exp(\lambda\cos\gamma_k).
\vspace*{0.1cm}
\end{equation}
Here, $\cos\gamma_k$ denotes the great-circle distance between points $\Omega$ and $\Omega_k$ on the sphere. The vMF distribution is a normal distribution on a sphere, centered around a mean direction $\Omega_k$ with a concentration parameter $\lambda$ -- the higher its value, the higher the concentration of the distribution around the mean direction (see Fig.~\ref{fig:A1} in Appendix~\ref{sec:vmf}). We write the normalization factor $1/R^2$ in Eq.~\eqref{eq:vmf} in analogy with the spherical expression of the Dirac $\delta$ function [Eq.~\eqref{eq:delta}].

Given $N$ charges on a sphere, the total surface charge distribution can thus be written as a sum of contributions from individual charges:
\begin{equation}
\label{eq:sig-vmf}
\sigma(\Omega)=\frac{e_0}{4\pi R^2}\sum_{k=1}^N\frac{q_k\lambda_k}{\sinh\lambda_k}\exp(\lambda_k\cos\gamma_k),
\end{equation}
where each charge is represented by its own vMF distribution characterized by the mean direction $\Omega_k$, coinciding with the position of the charge projected onto the unit sphere, and the charge's concentration parameter $\lambda_k$, describing its spatial extension around the mean position. The surface charge distribution, Eq.~\eqref{eq:sig-vmf}, can in turn be expanded in terms of its multipole moments
\begin{equation}
\label{eq:multipole}
\sigma(\Omega)=\frac{e_0}{4\pi R^2}\sum_{l,m}\sigma_{lm}Y_{lm}(\Omega).
\end{equation}
A lengthy derivation, given in Appendix~\ref{sec:derivation}, yields a very elegant expression for the multipole coefficients $\sigma_{lm}$,
\begin{equation}
\label{eq:slm}
\sigma_{lm}=4\pi\sum_kq_k\,g_l(\lambda_k)\,Y_{lm}^*(\Omega_k),
\end{equation}
where we have defined
\begin{equation}
\label{eq:glk}
g_l(\lambda)=\frac{\lambda}{\sinh\lambda}\,i_l(\lambda).
\end{equation}
Here, $i_l(x)$ are the modified spherical Bessel functions of the first kind~\cite{Abramowitz}. Rather unexpectedly, the multipole coefficients are determined by a single function dependent on the multipole order $\ell$ and the concentration parameter $\lambda_k$. With the knowledge of multipole coefficients [Eq.~\eqref{eq:slm}], we can now also insert them back into Eq.~\eqref{eq:sig-vmf} to obtain the total surface charge distribution.
%Setting all $g_l(\lambda_k)=1$, we obtain the coefficients of a multipole expansion of $\delta$ functions~\cite{ALB2017a}.

Given an expansion of a surface charge distribution in terms of its multipole coefficients, we define the multipole magnitude $S_l$ of order $\ell$ as
\begin{equation}
\label{eq:mag}
S_l=\sqrt{\frac{4\pi}{2l+1}\sum_m|\sigma_{lm}|^2}.
\end{equation}
Inserting the expression for the multipole coefficients, Eq.~\eqref{eq:slm}, we obtain the normalized multipole magnitudes (Appendix~\ref{sec:mags})
\begin{widetext}
\begin{equation}
\label{eq:Sl}
\frac{S_l}{|S_0|}=\left(\sum_k|q_k|\right)^{-1}\left[\sum_{k=t}q_k^2\,g_l^2(\lambda_k)+2\sum_{k>t}q_kq_t\,g_l(\lambda_k)\,g_l(\lambda_t)\,P_l(\cos\gamma_{kt})\right]^{1/2}.
\end{equation}
\end{widetext}
The monopole moment $S_0$ relates of course to the total charge $Q$, whereas the multipole moments of the first and second order correspond to the dipole and quadrupole moment, respectively, and can be easily related to their Cartesian forms~\cite{ALB2017a}. In order to enable an easy comparison between configurations with the same number of charges but different total charge, we have normalized the multipole magnitudes in Eq.~\eqref{eq:Sl} with the absolute value of the monopole moment, $|S_0|=4\pi\sum_k|q_k|=4\pi|Q|$.

We have thus derived the multipole coefficients for an arbitrary distribution of $N$ charges on a unit sphere [Eq.~\eqref{eq:slm}], where we have assigned to them vMF distributions with given mean directions $\Omega_k$ and concentration parameters $\lambda_k$. Through this, we have obtained a very simple expression both for the resulting total surface charge distribution and its corresponding multipole moments [Eq.~\eqref{eq:Sl}]. Such an approach ascribes a finite, continuous spatial extent to each charge, providing a more realistic description and at the same time avoiding the difficulties related to the multipole expansion of Dirac $\delta$ functions.

\subsection{Configurations of identical charges}

In order now to explore the consequences of the derived expressions for the surface charge distribution and its multipole moments, we will limit ourselves in the rest of the paper to configurations where all the charges possess identical properties, $q_k=q=1$ and $\lambda_k=\lambda$.

Such an assumption immediately enables us to study certain limiting cases of our results (Appendix~\ref{sec:limits}): When the concentration parameter $\lambda$ tends to $0$, the surface charge distribution of any configuration of charges expectedly becomes a uniform distribution on the sphere, described by its total charge. On the other hand, when $\lambda$ tends to infinity, the surface charge distribution reduces to a sum over Dirac $\delta$ functions of individual charges [Eq.~\eqref{eq:delta}].

More interestingly, the multipole magnitudes of a configuration of $N$ identical charges can be expressed as
\begin{eqnarray}
\frac{S_l}{|S_0|}&=&g_l(\lambda)\times\left(\frac{1}{N}+\frac{2}{N^2}\sum_{k>t}P_l(\cos\gamma_{kt})\right)^{1/2}\\
&=&g_l(\lambda)\times\frac{S_l^\infty}{|S_0|}.
\label{eq:sinf}
\end{eqnarray}
From here, we see that there are two major factors determining the relative contribution of a given multipole moment $S_l$ to the surface charge distribution. The first factor is given by the function $g_l(\lambda)$, and the second by the geometry of the configuration of the $N$ charges, given by their spherical distances $\cos\gamma_{kt}$. The latter are indeed all that determines the multipole magnitudes in the limit $\lambda\to\infty$, as $\lim_{\lambda\to\infty}g_l(\lambda)=1$ $\forall l$. On the other hand, when $\lambda\to0$, we have $\lim_{\lambda\to0}g_l(\lambda)\propto\lambda^l$ and the low-order multipoles become increasingly dominant. A more detailed discussion of the different limiting cases is given in Appendix~\ref{sec:limits}.

\section{Random configurations of identical charges and the role of symmetry}

We study the implications of our results by applying them to different configurations of $N$ identical charges on a unit sphere and analyzing the properties of the resulting surface charge distributions. All the charges share the same properties, $q_k=q=1$ and $\lambda_k=\lambda$, with $\lambda$ and $N$ being variable. For comparison, we use three different types of charge configurations on a sphere:
\begin{itemize}
\item [{\em (i)}] Random distributions of charges, where the positions of charges are distributed uniformly on the sphere.
\item [{\em (ii)}] Distributions of charges with some minimal distance between them. The positions of these charges are generated randomly and picked according to Mitchell's best-candidate algorithm~\cite{Mitchell1991} (approximating Poisson disc sampling and blue noise). In this manner, we prevent charges from being distributed too closely together, as can, for instance, happen in scheme {\em (i)}. The resulting positions of charges are, while random, no longer independent.
\item [{\em (iii)}] Distributions of charges based on the solutions of the Thomson problem, which minimizes the electrostatic energy of such a configuration~\cite{Wales2006}. Compared to schemes {\em (i)} and {\em (ii)}, the charges in these configurations are spaced the furthest apart, and the configurations exhibit various symmetries, including tetrahedral, octahedral, and icosahedral (depending on the number of charges $N$).
\end{itemize}
We will refer to these configurations as random, Mitchell, and Thomson configurations, respectively. While the solutions of the Thomson problem, {\em (iii)}, provide unique configurations for a given $N$, generating them randomly -- either uniformly, {\em (i)}, or with Mitchell's algorithm, {\em (ii)} -- can yield many different configurations. In the latter two cases we thus generate, for each $N$, $M=5000$ different configurations, allowing us to operate in terms of average quantities, where the average is taken over all $M$ configurations.

\begin{figure*}[!htp]
\includegraphics[width=1.5\columnwidth]{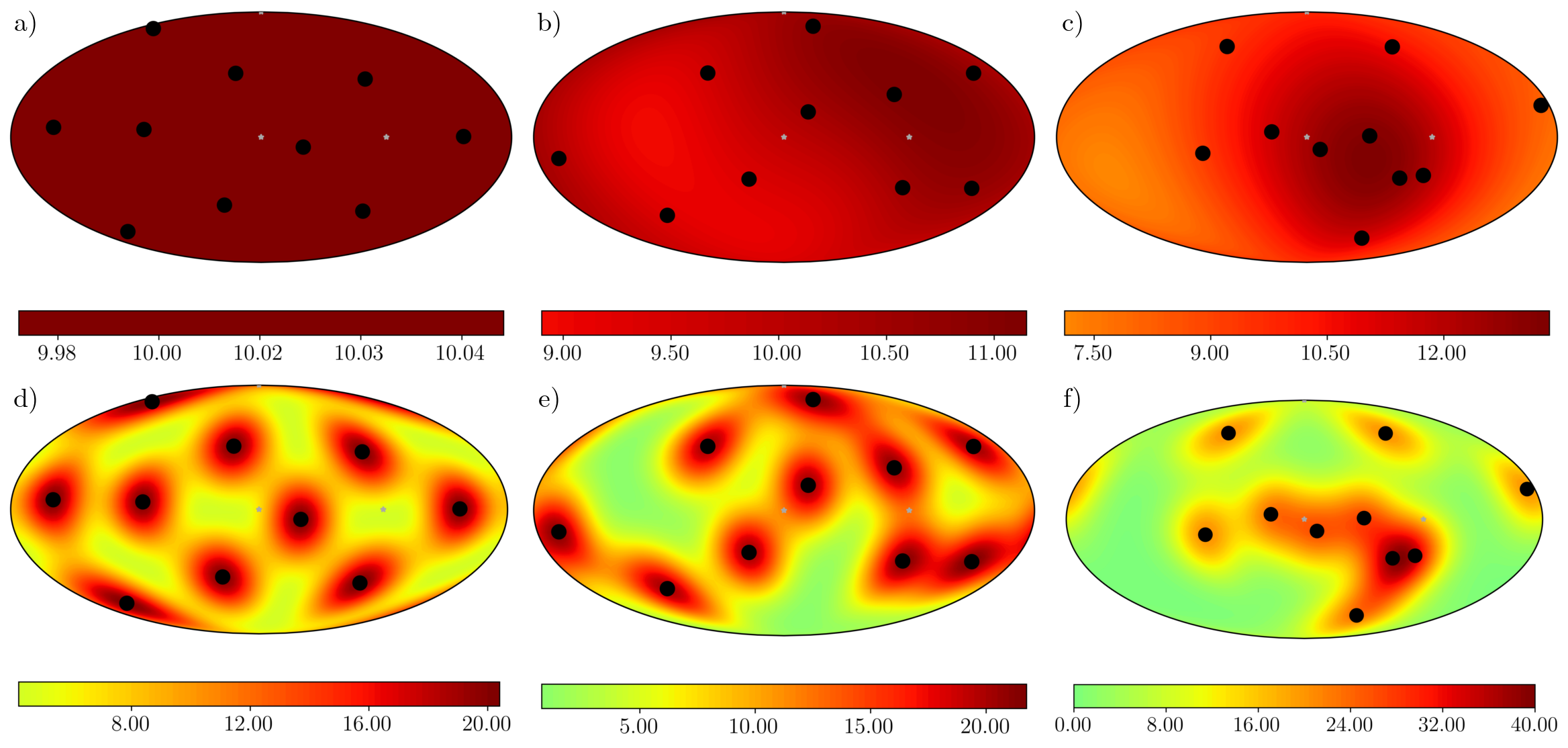}
\caption{Surface charge distributions for configurations of $N=10$ identical charges on a unit sphere. The distributions are mapped from a sphere onto a plane using Mollweide projection. {\bf (a)}, {\bf (d)} Configuration based on the solution of the Thomson problem; {\bf (b)}, {\bf (e)} configuration with charge positions generated using Mitchell's best candidate algorithm; {\bf (c)}, {\bf (f)} configuration generated by uniformly random positioning of charges. The value of the concentration parameter in panels {\bf (a)}-{\bf (c)} is $\lambda=1$, and in panels {\bf (d)}-{\bf (f)} $\lambda=10$. The color scheme shows the angular variation of the surface charge distribution in arbitrary units, with warmer colors indicating larger values.
\label{fig:1}}
\end{figure*}

In Fig.~\ref{fig:1}, we use the three different schemes {\em (i)}-{\em (iii)} to obtain configurations of $N=10$ identical charges and their surface charge distributions, which have been projected from a sphere onto a plane using Mollweide projection~\cite{Snyder}. In addition, the configurations are shown at two different values of the concentration parameter, $\lambda=1$ and $\lambda=10$ (cf.\ also Fig.~\ref{fig:A1} in Appendix~\ref{sec:vmf}). We see that, for small $\lambda$, the Thomson configuration is almost indistinguishable from a uniform charge distribution. Random and Mitchell configurations show more variation, especially if the charges are allowed to be located close to each other. At higher $\lambda$, where the influence of charges is more concentrated around their positions, the deviations from a uniform distribution become more prominent in all three configurations. Again, however, the relative positions of the charges determine the extent of this variation. These observations indicate that indeed $\lambda$ and the relative positions of the charges (given by $\cos\gamma_{kt}$) will determine the multipole characterization of a given configuration of charges. Mitchell's algorithm, {\em (ii)}, positions the charges so that there is a minimal distance between them, leading to a ``layered'' distribution of distances between charges; on the other hand, random positioning of charges onto the sphere, {\em (i)}, tends to distribute them uniformly on average (see Fig.~\ref{fig:E1} in Appendix~\ref{sec:extra}). Lastly, Thomson configurations exhibit the largest distances between particles and the highest overall symmetry.

\subsection{Multipole expansion}

Figure~\ref{fig:1} provides us with an insight into how a particular configuration of charges and their concentration parameter $\lambda$ influence the resulting surface charge distribution. However, as it is difficult to assess the general influence of the number of charges and their properties based on their surface charge distribution alone, we now turn our attention to their multipole magnitudes [Eq.~\eqref{eq:sinf}].

\begin{figure}[!htp]
\includegraphics[width=\columnwidth]{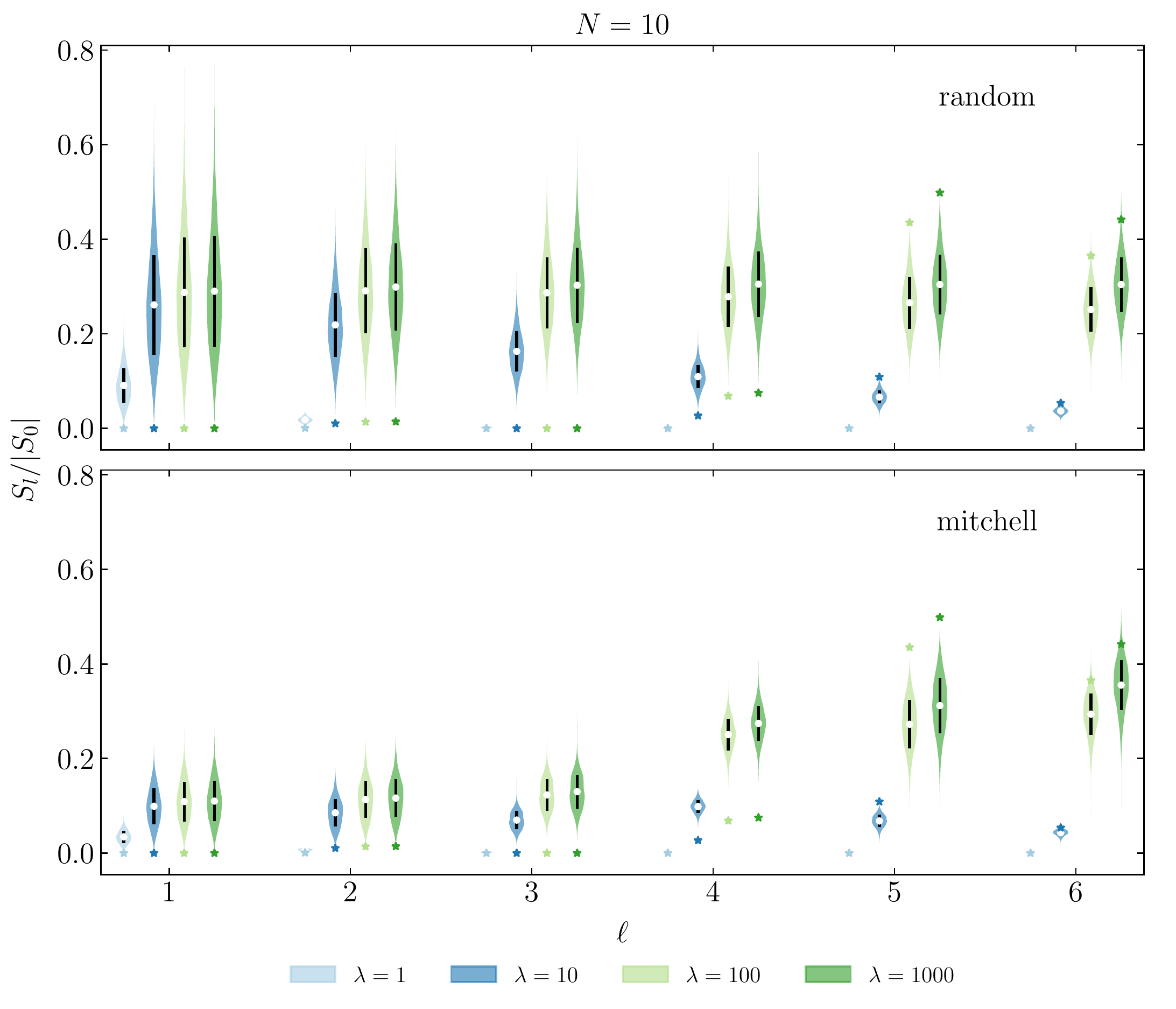}
\caption{Violin plot of the first $6$ multipole magnitudes for random and Mitchell configurations of $N=10$ identical charges. Each entry in the violin plot shows a (mirrored) distribution of normalized magnitudes of $5000$ different configurations, with the central symbols denoting the mean and the bars denoting the corresponding standard deviation. Star symbols show the multipole magnitudes of the corresponding Thomson configuration. The plot is shown for four different values of the concentration parameter $\lambda$.
\label{fig:2}}
\end{figure}

Figure~\ref{fig:2} shows the distributions of the first $6$ normalized multipole moments for $5000$ different random and Mitchell configurations of $N=10$ identical charges. We can see that, in both cases, at small values of $\lambda$ the multipole moments of high order $\ell$ are quickly suppressed, and the surface charge distribution is thus dominated by its monopole moment. When $\lambda$ increases, the high-order multipoles drop off ever more slowly until they become comparable among each other in the limit $\lambda\to\infty$. On average, random configurations tend to have much larger low-order multipoles (dipole, quadrupole) compared to Mitchell configurations with a minimum distance between the charges; this difference disappears for high-order multipoles. All these observations stem from the mean values of multipoles obtained by averaging over the $5000$ different configurations; within these, there is still a significant amount of variation, especially for low $\ell$.

\begin{figure}[!htp]
\includegraphics[width=\columnwidth]{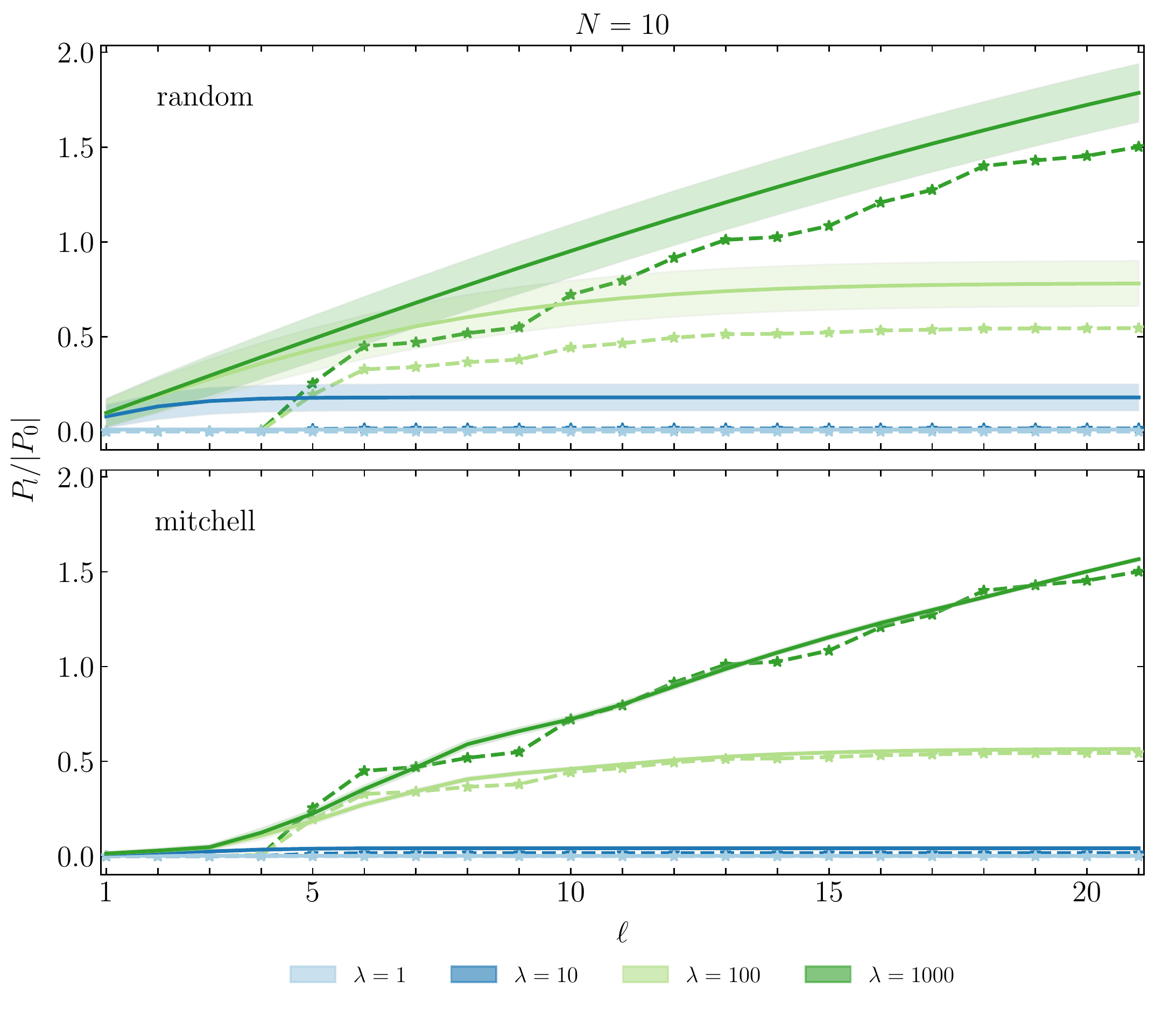}
\caption{Normalized total power $P_l/|P_0|$ of random and Mitchell configurations of $N=10$ identical charges, obtained by summing the squares of multipole magnitudes up to  order $\ell$. Full lines show the mean values, obtained by averaging over $5000$ different configurations, while the shaded regions denote the corresponding standard deviations. The latter are negligible for Mitchell configurations. Dashed lines and star symbols show the total power for the corresponding Thomson configuration. The plot is shown for four different values of the concentration parameter $\lambda$.
\label{fig:3}}
\end{figure}

As the concentration parameter $\lambda$ is increased, multipoles of high order become less and less negligible (Fig.~\ref{fig:2}). This behavior becomes even more pronounced when we plot the normalized total power of order $\ell$, $P_l/|P_0|$, obtained by terminating the expression for total power [Eq.~\eqref{eq:pow}] at a given $\ell$. The total power consists of a sum of squared multipole magnitudes, and is shown in Fig.~\ref{fig:3} for configurations of $N=10$ identical charges. Again, we can see clearly that at low $\lambda$, the surface charge distribution of any configuration is dominated by its monopole moment. Upon a gradual increase in $\lambda$, the next few multipole moments become more important, while the majority of the multipoles still do not contribute to the total power. However, with a still further increase in $\lambda$, more and more multipoles need to be summed before the total power converges, and when $\lambda=1000$, we are far from convergence even when we truncate the sum only at $\ell=21$. This observation holds regardless of which of the three configuration schemes {\em (i)}-{\em (iii)} we choose. We can also observe that the total power of Mitchell configurations of charges, keeping a minimum distance between them, matches quite closely the total power of the corresponding Thomson configuration, while the total power of random configurations is always higher, especially due to the higher values of dipole and quadrupole moments.

Figures~\ref{fig:E2} and~\ref{fig:E3} in Appendix~\ref{sec:extra} show the results for configurations of $N=20$ identical charges, analogous to those presented in Figs.~\ref{fig:2} and~\ref{fig:3} for configurations of $N=10$ charges. We can see that the general behavior is similar in the two cases. Notably, though, a higher number of charges lowers the overall magnitudes of multipoles compared to the monopole moment, and in case of Mitchell configurations, the first non-negligible multipole occurs at a later value of $\ell$ compared to the case of $N=10$ charges ($\ell\sim8$ and $\ell\sim6$, respectively).

\begin{figure*}[!htp]
\includegraphics[width=2\columnwidth]{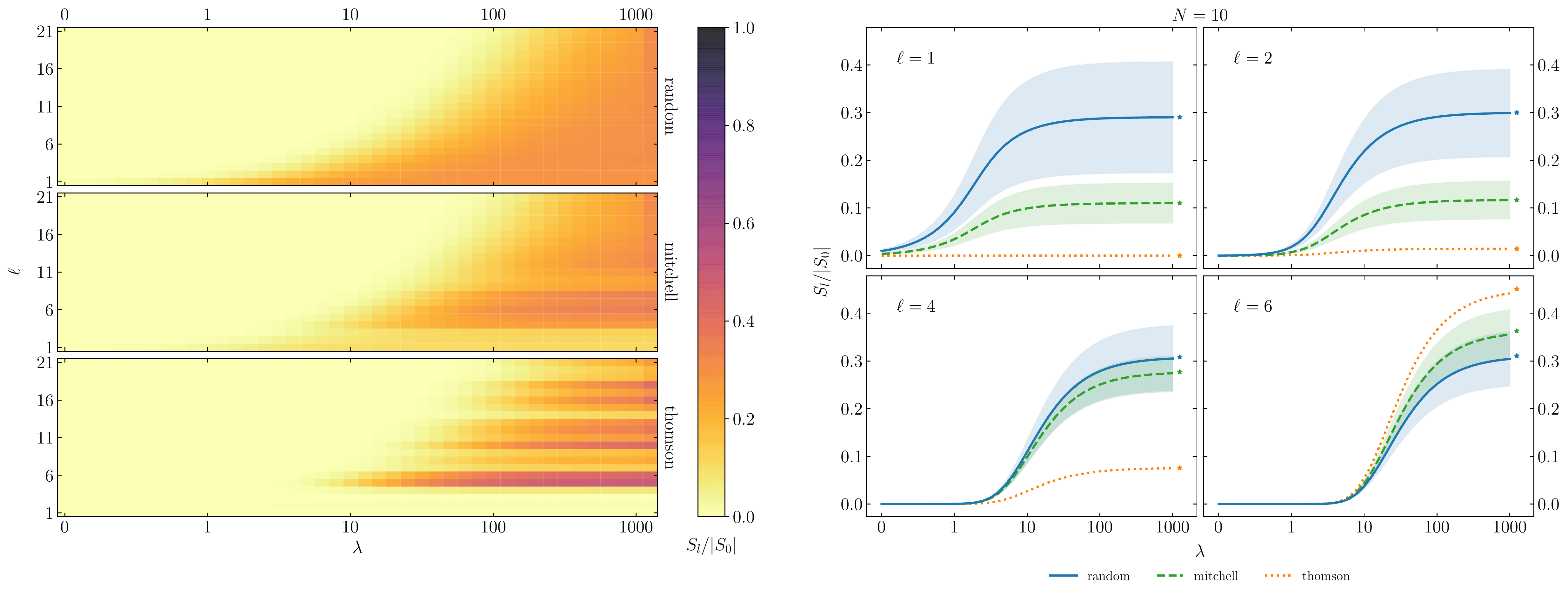}
\caption{Heatmap of normalized multipole magnitudes $S_l/|S_0|$ as a function of $\lambda$ and $\ell$ for different configurations of $N=10$ identical charges. Plots isolate the $\lambda$-dependence of multipoles with order $\ell=1$, $2$, $4$, and $6$. In the case of random and Mitchell configurations, the heatmap and the full lines in the plots show the mean values obtained by averaging over $5000$ different configurations, while the shaded regions denote the corresponding standard deviations. The last column of the heatmap and the star symbols in the plots show the values of multipole magnitudes in the limit $\lambda\to\infty$, $S_l^\infty/|S_0|$.
\label{fig:4}}
\end{figure*}

The first non-negligible multipole in an expansion of a surface charge distribution thus appears to be in large part determined by any symmetry a given configuration might possess. Thomson configuration of $N=10$ identical charges possesses a $D_{4d}$ symmetry~\cite{Wales2006}, and as such, its first non-vanishing multipole should be the quadrupole, $\ell=2$~\cite{Gelessus1995}. From Fig.~\ref{fig:4} we can observe that this is indeed the case. Multipole magnitudes reach a first peak, however, at $\ell=5$ and $\ell=6$. Compared with the corresponding random and Mitchell configurations, the Thomson configuration also shows the most variation between multipoles of different order -- that is, while some multipoles are strongly represented in the expansion, others are completely absent; this is especially noticeable in the limit $\lambda\to\infty$, where the only contribution to the multipoles is due to the geometry of the configuration [Eq.~\eqref{eq:sinf}]. We can see that the higher the order of the multipole, $\ell$, the more slowly this limiting value is reached.

The results for configurations of $N=10$ identical charges indicate that the multipole description of a surface charge distribution and its deviations from a uniform distribution are influenced by several factors (Fig.~\ref{fig:4}). In configurations where high-order multipoles (large $\ell$) are dominant, the  description of charges will approach the limit of Dirac $\delta$ functions ($\lambda\to\infty$) slowly, i.e., their surface charge distribution will be approximated well by a uniform distribution in a wider range of $\lambda$s. Another factor influencing the deviation from a uniform distribution is the limiting value of the multipole magnitudes, $S_l^\infty$, dependent solely on the geometrical distribution of the charges. A lower limiting value, typical of completely random distributions, implies that, in spite of how quickly this limit is attained with $\lambda$, the uniform distribution given by the monopole moment will remain dominant.

\begin{figure}[!htp]
\includegraphics[width=\columnwidth]{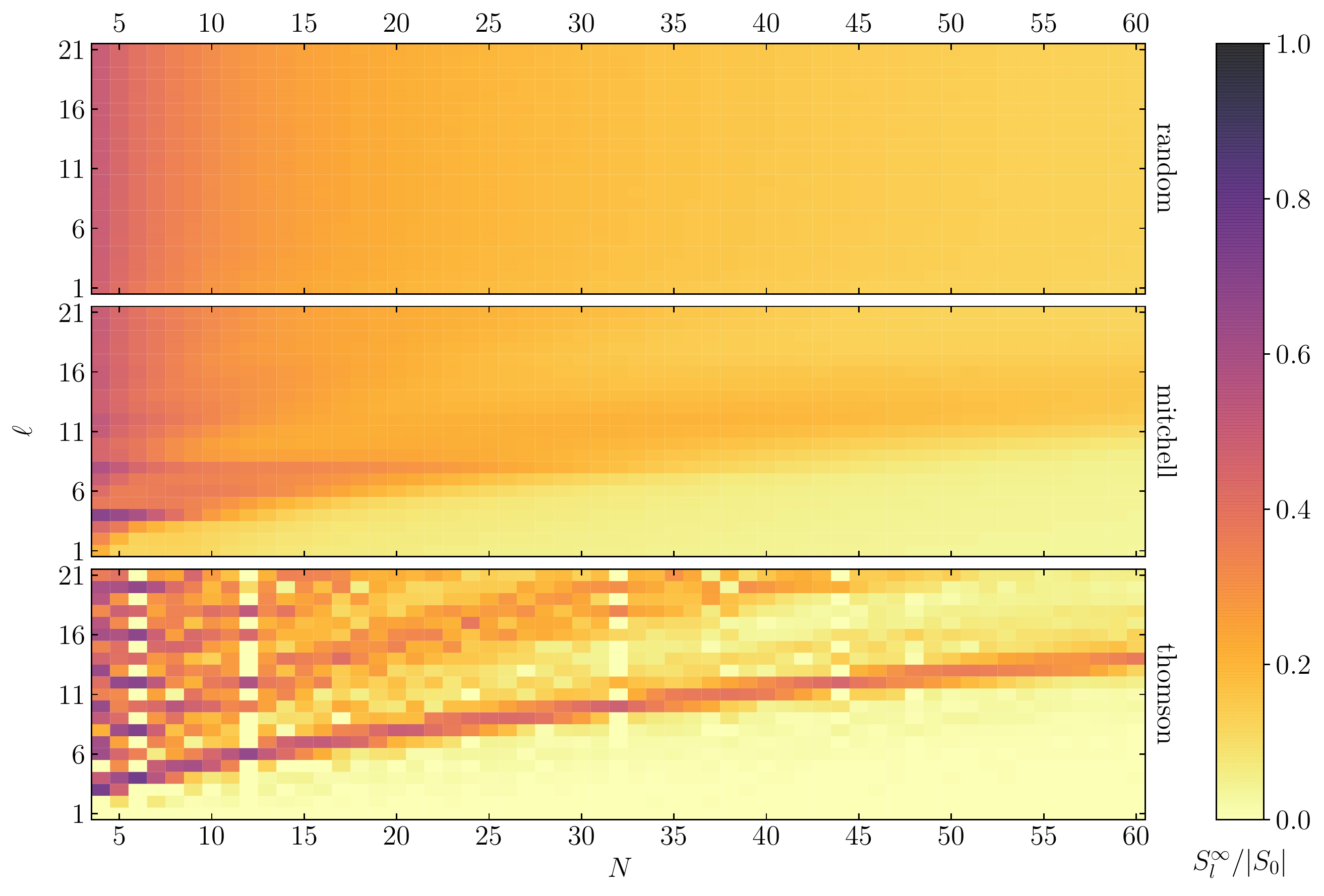}
\caption{Heatmap showing the limiting value of the multipole magnitudes $S_l^\infty/|S_0|$ as a function of $N$ and $\ell$. The value of $S_l^\infty/|S_0|$ is obtained in the limit $\lambda\to\infty$, and depends solely on the distances between charges in a given configuration. In the case of random and Mitchell configurations, the heatmap shows the mean values obtained by averaging over $5000$ different configurations.
\label{fig:5}}
\end{figure}

\begin{figure}[!htp]
\includegraphics[width=\columnwidth]{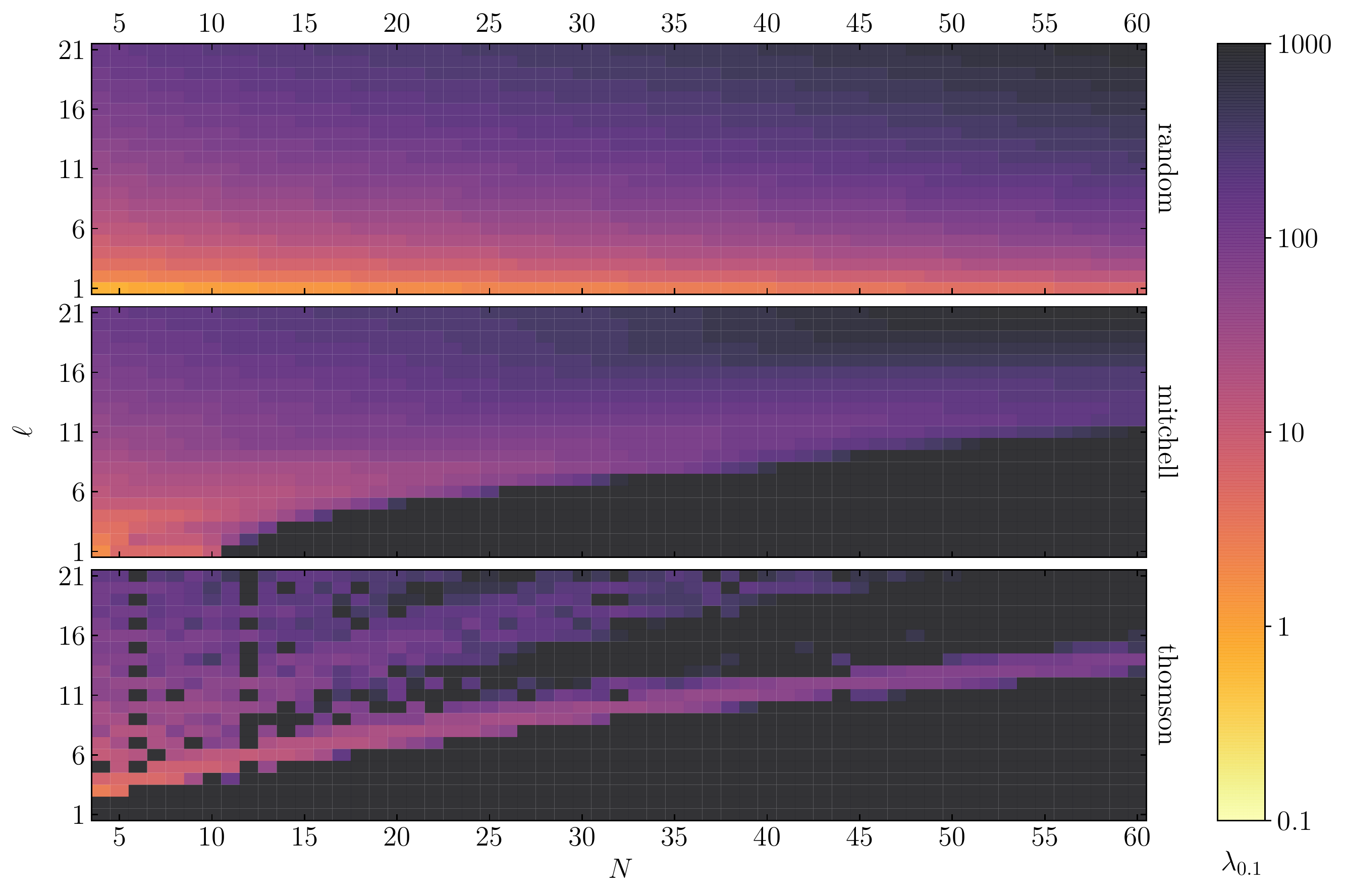}
\caption{Heatmap showing the value of the concentration parameter where a multipole magnitude reaches $10\%$ of the monopole magnitude (total charge), $\lambda_{0.1}$, as a function of $N$ and $\ell$. A value of $\lambda_{0.1}=1000$ indicates that a multipole does not reach the $10\%$ value of the monopole in the considered range of $\lambda$s. In the case of random and Mitchell configurations, the heatmap shows the mean values obtained by averaging over $5000$ different configurations.
\label{fig:6}}
\end{figure}

We now wish to generalize these observations to configurations with an arbitrary number of charges. In order to do that, we will characterize the behavior of multipole magnitudes $S_l$ with two parameters: first, with their limiting value $S_l^\infty$, and second, with the value of the concentration parameter where a multipole magnitude reaches $10\%$ of the monopole moment, which we will denote $\lambda_{0.1}$:
\begin{equation}
\left.\frac{S_l}{|S_0|}\right|_{\lambda_{0.1}}=g_l(\lambda_{0.1})\times\frac{S_l^\infty}{|S_0|}=0.1.
\end{equation}
The dependence of these two parameters on the number of identical charges in a configuration, $N$, is shown in Figs.~\ref{fig:5} and~\ref{fig:6} for a large number of multipole magnitudes. First of all, we see that, for random configurations of charges, the limiting value $S_l^\infty$ changes only slowly with $\ell$, while it decreases with $N$ (Fig.~\ref{fig:5}). However, the value of $\ell$ influences rather strongly the parameter $\lambda_{0.1}$ -- the speed at which the limiting value $S_l^\infty$ is attained (Fig.~\ref{fig:6}). In contrast, $\lambda_{0.1}$ is not influenced much by the number of charges in a random configuration.

The number of charges $N$ has a much bigger influence on the behavior of Thomson configurations. Partially, this is to be expected, as they exhibit various symmetries at different $N$, resulting in a strong presence of multipoles of given $\ell$ and on the almost complete vanishing of other multipoles. For instance, it is known that for configurations with tetrahedral, octahedral, and icosahedral symmetries, only certain values of $\ell$ are permitted in the multipole expansion of the surface charge distribution~\cite{Lorman2007,ALB2013a}: $\ell_\mathrm{tet}=4i+6j\,(+3)$, $\ell_\mathrm{oct}=4i+6j\,(+9)$, and $\ell_\mathrm{ico}=6i+10j\,(+15)$; configurations with odd values of $\ell$ are those which lack inversion symmetry. The symmetries of different configurations and the permitted multipoles in their expansion are reflected in a checkered pattern in the heatmaps in Figs.~\ref{fig:5} and~\ref{fig:6}, a pattern which is completely absent in the case of random configurations. 

In addition to this, Thomson configurations exhibit vanishing multipole moments of low order $\ell$, the amount of which increases with increasing $N$. For example, while the dipole and quadrupole moment are negligible for a Thomson configuration with $N=10$ charges, all of the first $10$ multipole moments are negligible for a Thomson configuration with $N=60$ charges. What is more, these vanishing multipole moments appear to occur periodically in ``islands'' of $\ell$ numbers, the extent of which increases with $N$ (see also Fig.~\ref{fig:E4} in Appendix~\ref{sec:extra}).

Mitchell configurations present a middle ground between random and Thomson configurations. The variation of $S_l^\infty$ and $\lambda_{0.1}$ occurs gradually with $N$, in a similar fashion to random configurations, yet at the same time we can observe the same vanishing multipoles of low order at higher numbers of charges as we saw for Thomson configurations. Individual multipoles also tend to be more pronounced in Mitchell configurations compared to the multipoles of random configurations, yet not so strikingly as in the symmetric Thomson configurations.

Taken together, our results for configurations of identical charges show that the multipole magnitudes of their surface charge distributions depend strongly on the exact geometry of the configuration, with clear differences between randomly positioned charges, configurations where charges are placed a minimum distance apart, and configurations of very high symmetry. In addition, both the number of charges and their size -- given by the concentration parameter -- place the surface charge distribution of a configuration of charges in different regimes, where the distribution can either behave solely as a uniform distribution, or needs a large number of high-order multipoles to be accurately represented.

\section{Discussion and conclusions}

In this work, we have presented a novel way of constructing continuous surface charge distributions of spherical particles composed of numerous charges. Our approach is based on the description of individual charges with a vMF distribution on a sphere, taking into account the finite extent of the charges. With this, we were able to extract the electrostatic multipoles of such surface charge distributions and analyze their behavior as a function of the multipole order $\ell$, and the number $N$ and size (concentration parameter $\lambda$) of the charges. Analytically, we have derived the precise relation of the multipole magnitudes to the size of the charges and the geometry of their configuration on the sphere. We have explored the predictions of our approach on different configurations of identical charges, generated either randomly or by using Mitchell's algorithm, or extracted from the solutions of the Thomson problem.

While we have considered configurations of charges with identical properties, the results derived in this paper allow an easy generalization to arbitrary configurations of fractional charges $q_k$ with concentration parameters $\lambda_k$ (which take into account their relative extension on a unit sphere with $R=1$). In addition, given a ``physical'' size of a charge $a_k$, we can rewrite the concentration parameter for an arbitrary size of the sphere $R$ as
\begin{equation}
\label{eq:elk}
\lambda_k=\frac{R}{a_k}.
\end{equation}
This implies that, for a given size of a charge, the parameter $\lambda$ will be larger for larger spheres, where the same charge will appear more localized on a larger than on a smaller sphere. In this way, our approach can be used to study the surface charge distributions on biomolecules of different sizes, ranging from small globular proteins ($R\gtrsim1$~nm) to larger viral capsids ($R\gtrsim10$-$20$ nm)~\cite{ALB2017a,ALB2013b}, thus spanning a range of $\lambda_k\gtrsim1$-$100$, depending on the size of the macromolecule in question. This of course makes it necessary to be able to estimate the value of the parameter $a_k$, which can be obtained from the biochemical nature of the molecules (such as different amino acids) carrying the charge in a given system.

An important conclusion we can draw from our results is that the relationship between the size of the charges relative to the size of the sphere they are located on plays a significant role in determining the resulting surface charge distributions. We have seen that going from very spread-out charges (small $\lambda$) to charges that can be treated as Dirac $\delta$ functions (large $\lambda$) results in a wildly different relative importance of the corresponding multipole magnitudes. Specifically, a surface charge distribution constructed out of point charges will need in principle an infinite sum of multipoles in order to be represented accurately, potentially masking the importance of low-order multipoles. Consequently, such a description could lead to an over- or underestimation of dipole and quadrupole moments. Our results also indicate that even in descriptions of general charge distributions of molecules, taking into account the finite extension of charges could have a pronounced effect on the determination of their electrostatic multipoles~\cite{Stone1981,Larsson1985}.

In general, our approach also helps distinguish the regime where a given configuration of charges on a sphere can be described well by a uniform distribution from the regime where the charges are localized enough that their geometry and symmetry determine the largest multipoles in the expansion of the surface charge distribution. While the geometry of a particular configuration of charges turns out to play a large role, it will nonetheless tend to a uniform distribution when $\lambda\ll1$, whereas the multipole magnitudes will be determined solely by the geometry of the configuration when $\lambda\gg1$. In the intermediate regime of $\lambda$s, increasing the number of charges in a configuration will in general reduce the importance of high-order multipoles, the more so the less symmetric the configuration. At the same time, multipoles of low order are prominent at small values of $\lambda$, and the high-order multipoles become comparable only when $\lambda$ is increased.

The parameter space of biological macromolecules and colloids can in fact span a large range of values studied in this work. The charge of both small globular proteins and large capsid assemblies is carried by the same amino acids, meaning that the concentration parameters of charges will be smaller for the globular proteins than for viral capsids. On the other hand, viral capsids can carry several hundreds or thousands of individual charges, while the smaller proteins are often composed of only a few tens of charges. Consequently, we can expect a large variation in the multipole behavior of the surface charge distributions in different systems.

A particular observation that should be of importance when describing the surface charge distributions in viral capsids is that the order $\ell$ of the dominant multipole in symmetric distributions increases with an increasing number of charges in a configuration. This is in contrast to random configurations of charges, which tend to be dominated by low-order multipoles, no matter the number of charges. As viral capsids possess very high symmetry -- typically icosahedral -- our approach can be used to extract the dominant multipole describing the symmetry of a particular configuration, which was recently shown to play a role in orientational phase transitions in capsids~\cite{Dharmavaram2017}.

The approach presented in this work enables a simple yet powerful construction of continuous surface charge distributions from individual charges on spherical particles, taking into account the finite size of each charge. This allows for a construction of various analytical models based on multipole expansion that can be used in describing systems of inverse patchy colloids, small globular proteins, and viral capsids of different sizes. In addition, the approach presented here can help elucidate the relative relevance of multipole magnitudes in a given system, and can help distinguish the cases where total charge provides a sufficient description of the electrostatic properties from the cases where a more detailed multipole expansion is needed.

\acknowledgments

I thank S.\ \v{C}opar and R.\ Podgornik for numerous helpful discussions and comments. This work was supported by the Slovenian Research Agency (under Research Core Funding grant No. P1-0055).

\appendix

\section{\label{sec:vmf} von Mises-Fisher distribution}

Von Mises-Fisher (vMF) distribution is a normal probability distribution on the $(p-1)$-dimensional sphere in $\mathcal{R}^p$~\cite{Mardia2009}. The vMF distribution for a random $p$-dimensional vector $\mathbf{r}$ is given by
\begin{equation}
f_p(\mathbf{r}\,|\,\mathbf{r}_0,\lambda)=C_p(\lambda)\,\exp(\lambda\,\mathbf{r}_0^T\mathbf{r}).
\end{equation}
Here, $\lambda\geqslant0$, $|\mathbf{r_0}|$=1, and the normalization constant $C_p$ is equal to
\begin{equation}
C_p(\lambda)=\frac{\lambda^{p/2-1}}{(2\pi)^{p/2}I_{p/2-1}(\lambda)},
\end{equation}
where $I_\nu$ denotes the modified Bessel functions of the first kind~\cite{Abramowitz}. vMF distribution is a normal distribution on a sphere, where the parameter $\mathbf{r}_0$ is the mean direction of the distribution, and the parameter $\lambda$ is the concentration parameter -- the higher its value, the higher the concentration of the distribution around the mean direction. A generalization of the vMF distribution to a bivariate normal distribution with an unconstrained covariance matrix is called the spherical Fisher-Bingham or Kent distribution~\cite{Kent1982}.

\begin{figure*}[!htp]
\includegraphics[width=1.5\columnwidth]{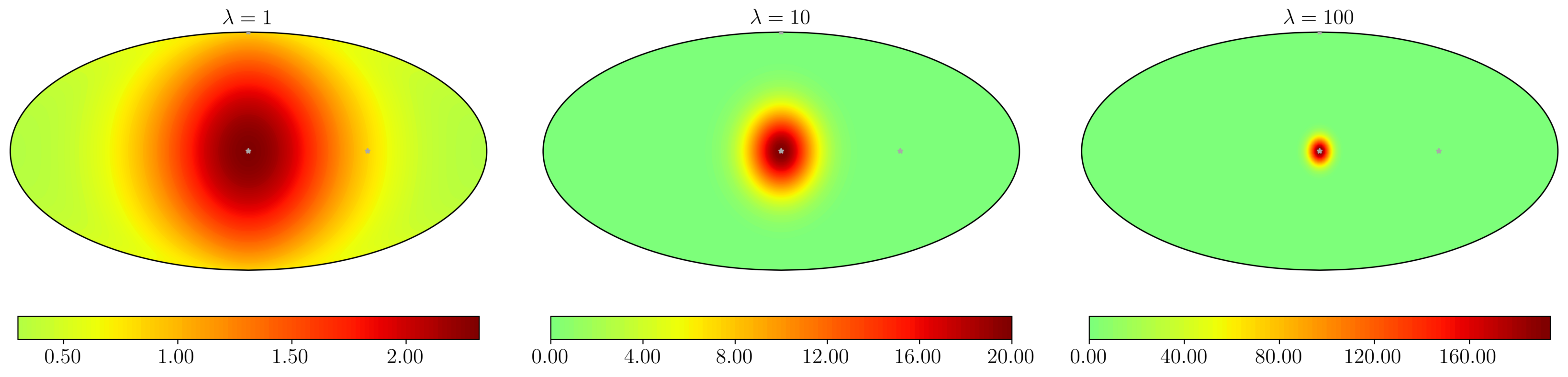}
\caption{Distribution of a single point charge with $q=1$ located on the $x$ axis on a unit sphere. The distribution is shown for three different values of $\lambda=1$, $10$, $100$. The distribution was mapped from a sphere to a plane using Mollweide projection, and the coordinate axes are shown with gray stars.
\label{fig:A1}}
\end{figure*}

In three dimensions -- on a unit sphere $\mathcal{S}_2$ -- the normalization constant of the vMF distribution reduces to
\begin{equation}
C_3=\frac{\lambda}{4\pi\sinh\lambda}=\frac{\lambda}{4\pi(e^\lambda-e^{-\lambda})},
\end{equation}
and we can thus write
\begin{equation}
f_3(\mathbf{r}\,|\,\mathbf{r}_0,\lambda)=\frac{\lambda}{4\pi\sinh\lambda}\,\exp(\lambda\,\mathbf{r}_0^T\mathbf{r}).
\end{equation}
Since any vector on the unit sphere can be represented in spherical coordinates as
\begin{equation}
\mathbf{r}=(\cos\varphi\sin\vartheta,\sin\varphi\sin\vartheta,\cos\vartheta).
\end{equation}
the exponent of the vMF distribution becomes
\begin{equation}
\exp(\lambda\,\mathbf{r}_0^T\mathbf{r})=\exp(\lambda\cos\gamma_0),
\end{equation}
where $\gamma_0$ denotes the great-circle distance between points $\Omega$ and $\Omega_0$,
\begin{equation}
\cos\gamma_0=\cos\vartheta\cos\vartheta_0+\cos(\varphi-\varphi_0)\sin\vartheta\sin\vartheta_0.
\end{equation}
With this, we can write the vMF distribution on a unit sphere centered around a point $\Omega_0$ as
\begin{equation}
f_3(\Omega\,|\,\Omega_0,\lambda)=\frac{\lambda}{4\pi\sinh\lambda}\exp(\lambda\cos\gamma_0).
\end{equation}
The distribution is normalized so that
\begin{equation}
\oint_{\mathcal{S}_2}\mathrm{d}\Omega\,f_3(\Omega\,|\,\Omega_0,\lambda)=1.
\end{equation}
The parameter $\lambda$ determines the concentration of the distribution centered around $\Omega_0$. For $\lambda=0$, the distribution is uniform on the sphere, while for $\lambda\to\infty$, the distribution tends to a Dirac $\delta$ function. Applying this to a distribution of a single point charge on a unit sphere [Eq.~\eqref{eq:delta}], Fig.~\ref{fig:A1} shows the distribution of a charge with $q=1$ located on the $x$-axis, for three different values of $\lambda$. When the concentration parameter is small, $\lambda=1$, the distribution extends across most of the unit sphere; however, with increasing $\lambda$, the influence of the charge becomes more and more localized.

\section{\label{sec:derivation} Derivation of multipole coefficients of the vMF surface charge distribution}

We start with the vMF surface charge distribution of a number of point charges $q_k$ with concentration parameters $\lambda_k$ and centered on positions $\Omega_k$ [Eq.~\eqref{eq:sig-vmf}]. From the corresponding multipole expansion, Eq.~\eqref{eq:multipole}, we obtain for the multipole coefficients
\begin{equation}
\sigma_{lm}=\sum_k\frac{q_k\lambda_k}{\sinh\lambda_k}\oint\mathrm{d}\Omega\,Y_{lm}^*(\Omega)\,\exp(\lambda_k\cos\gamma_k).
\end{equation}
The exponential function can be written as a power series, wherefrom we get
\begin{equation}
\label{eq:tmp}
\sigma_{lm}=\sum_k\frac{q_k\lambda_k}{\sinh\lambda_k}\sum_n\frac{\lambda_k^n}{n!}\oint\mathrm{d}\Omega\,Y_{lm}^*(\Omega)\,\cos^n\gamma_k.
\end{equation}
Introducing $x_k=\cos\gamma_k$, we split the sum over $n$ into even and odd terms:
\begin{eqnarray}
\label{eq:split}
\nonumber\sigma_{lm}&=&\sum_k\frac{q_k\lambda_k}{\sinh\lambda_k}\left\{\oint\mathrm{d}\Omega\,Y_{lm}^*(\Omega)\sum_n\frac{\lambda_k^{2n}}{(2n)!}x_k^{2n}\right.\\
&&+\left.\oint\mathrm{d}\Omega\,Y_{lm}^*(\Omega)\sum_n\frac{\lambda_k^{2n+1}}{(2n+1)!}x_k^{2n+1}\right\}.
\end{eqnarray}
Based on Ref.~\cite{Arfken}, we postulate that
\begin{equation}
\label{eq:lemma}
\sum_{n=0}^{\infty}\alpha_nx_k^n=\sum_{m=0}^{\infty}a_mP_m(x_k),
\end{equation}
where $P_n(x)$ are the Legendre polynomials, and the sum runs either over $m=n=\mathrm{even}$ or $m=n=\mathrm{odd}$. Using the orthogonality of spherical harmonics, we can write
\begin{equation}
\sum_n\alpha_n\int_{-1}^1x^nP_m(x)\mathrm{d}x=\frac{2}{2m+1}\,a_m.
\end{equation}
The integral can be split in two parts, and taking into account $P_m(-x)=(-1)^mP_m(x)$, we see that
\begin{widetext}
\begin{equation}
\label{eq:am}
a_m=\frac{2m+1}{2}\sum_n\alpha_n\times\begin{dcases}
2\int_0^1x^nP_m(x)\mathrm{d}x & \text{,\quad$n+m$ even}\\
0 & \text{,\quad$n+m$ odd}
\end{dcases}\quad,
\end{equation}
and the integral can be expressed in terms of $\Gamma$ functions~\cite{Abramowitz}:
\begin{equation}
\int_0^1x^nP_m(x)\mathrm{d}x=\frac{\sqrt{\pi}\,2^{-n-1}\Gamma(1+n)}{\Gamma(1+n/2-m/2)\Gamma(3/2+n/2+m/2)}.
\end{equation}
By writing $a_m=\sum_n A_{mn}\alpha_n$, we get from Eq.~\eqref{eq:am}
\begin{equation}
A_{mn}=\frac{\sqrt{\pi}\,2^{-n-1}\,(2m+1)\Gamma(1+n)}{\Gamma(1+n/2-m/2)\Gamma(3/2+n/2+m/2)}\quad,\quad n+m\textrm{ even;}\quad m\leqslant n,
\end{equation}
and $0$ otherwise. We immediately see that when $n=\mathrm{even}$, so is $m$, and conversely, when $n=\mathrm{odd}$, so is again $m$. Summing over even powers of $x$ in Eq.~\eqref{eq:lemma} will thus yield only $P_n(x)$ of even order, and similarly for the sum over odd powers. In addition, the coefficients $A_{mn}$ are nonzero only when $m\leqslant n$.
From Eq.~\eqref{eq:split} we have $\alpha_n=\lambda^n/n!$, and so it follows that
\begin{equation}
\label{eq:am2}
a_m=\sum_{n\geqslant m}^\infty A_{mn}\,\frac{\lambda^n}{n!}=\sum_{n\geqslant m}^\infty\frac{\lambda^n\sqrt{\pi}\,2^{-n-1}\,(2m+1)}{\Gamma(1+n/2-m/2)\,\Gamma(3/2+n/2+m/2)}\quad,\quad n+m\textrm{ even.}
\end{equation}
Inserting now the theorem in Eq.~\eqref{eq:lemma} into Eq.~\eqref{eq:split}, we obtain
\begin{equation}
\sigma_{lm}=\sum_k\frac{q_k\lambda_k}{\sinh\lambda_k}\left\{\oint\mathrm{d}\Omega\,Y_{lm}^*(\Omega)\sum_sa_{2s}P_{2s}(x_k)+\oint\mathrm{d}\Omega\,Y_{lm}^*(\Omega)\sum_sa_{2s+1}P_{2s+1}(x_k)\right\}.
\end{equation}
Next, we use the addition theorem for the spherical harmonics to write
%\begin{widetext}
\begin{eqnarray}
\nonumber\sigma_{lm}&=&\sum_k\frac{q_k\lambda_k}{\sinh\lambda_k}\left\{\sum_sa_{2s}\oint\mathrm{d}\Omega\,Y_{lm}^*(\Omega)\,\frac{4\pi}{2(2s)+1}\sum_{t=-2s}^{2s}Y_{2s,t}(\Omega)\,Y_{2s,t}^*(\Omega_k)+\right.\\
&&\phantom{\sum_k\frac{q_k\lambda_k}{\sinh\lambda_k}\left\{\right.}\left.\sum_sa_{2s+1}\oint\mathrm{d}\Omega\,Y_{lm}^*(\Omega)\,\frac{4\pi}{2(2s+1)+1}\sum_{t=-2s+1}^{2s+1}Y_{2s+1,t}(\Omega)\,Y_{2s+1,t}^*(\Omega_k)\right\}.
\end{eqnarray}
%\end{widetext}
Rearranging the order of summation and integration, the integrals evaluate into Dirac $\delta$ functions. By virtue of this, the sums over $s$ and $t$ disappear, yielding
\begin{equation}
\label{eq:slm1}
\sigma_{lm}=\sum_k\frac{q_k\lambda_k}{\sinh\lambda_k}\times\begin{dcases}
a_l\,\frac{4\pi}{2l+1}\,Y_{l,m}^*(\Omega_k) & \text{,\quad$l$ even}\\
a_l\,\frac{4\pi}{2l+1}\,Y_{l,m}^*(\Omega_k) & \text{,\quad$l$ odd}
\end{dcases}\quad.
\end{equation}
Using the expression for the coefficients $a_m$ in Eq.~\eqref{eq:am2}, we obtain from Eq.~\eqref{eq:slm1}:
\begin{equation}
\label{eq:slm2}
\sigma_{lm}=4\pi\sum_kq_k\,Y_{lm}^*(\Omega_k)\sum_{s\geqslant l}^\infty\frac{\lambda^{s+1}_k}{\sinh\lambda_k}\,\frac{\sqrt{\pi}\,2^{-s-1}}{\Gamma(s/2-l/2+1)\,\Gamma(s/2+l/2+3/2)}\quad,\quad l+s\textrm{ even},
\end{equation}
which holds true for both even and odd $l$. Introducing
\begin{equation}
g_l(\lambda)=\sum_{s\geqslant l}^\infty\frac{\lambda^{s+1}}{\sinh\lambda}\,\frac{\sqrt{\pi}\,2^{-s-1}}{\Gamma(s/2-l/2+1)\,\Gamma(s/2+l/2+3/2)}\quad,\quad l+s\textrm{ even},
\end{equation}
\end{widetext}
we can write the multipole coefficients as
\begin{equation}
\label{eq:slm3}
\sigma_{lm}=4\pi\sum_kq_k\,g_l(\lambda_k)\,Y_{lm}^*(\Omega_k).
\end{equation}
What is more, using Mathematica software~\cite{Mathematica} we can show that the function $g_l(\lambda)$ evaluates to
\begin{equation}
g_l(\lambda)=\frac{\lambda}{\sinh\lambda}\,i_l(\lambda),
\end{equation}
where
\begin{equation}
i_l(x)=\sqrt{\frac{\pi}{2x}}\,I_{l+1/2}(x)
\end{equation}
are the modified spherical Bessel functions of the first kind~\cite{Abramowitz}. Thus, we finally obtain the result of Eq.~\eqref{eq:slm}.

\section{\label{sec:mags}Multipole magnitudes, total power, and bond order parameters}

In order to obtain the multipole magnitudes from Eq.~\eqref{eq:mag}, we insert the expression for the multipole coefficients, Eq.~\eqref{eq:slm}, into the squared form of the magnitudes. Thus, we get
\begin{eqnarray}
\nonumber S_l^2&=&\frac{(4\pi)^3}{2l+1}\sum_m\left[\left(\sum_kq_k\,g_l(\lambda_k)\,Y_{lm}^*(\Omega_k)\right)\right.\\
&&\times\left.\left(\sum_tq_t\,g_l(\lambda_t)\,Y_{lm}(\Omega_t)\right)\right].
\end{eqnarray}
Using the addition theorem for spherical harmonics, we then obtain
\begin{equation}
S_l^2=(4\pi)^2\sum_{k,t}q_k\,q_t\,g_l(\lambda_k)\,g_l(\lambda_t)\,P_l(\cos\gamma_{kt}).
\end{equation}
Taking into account that $P_l(\cos\gamma_{kk}=1)=1$ and that $\cos\gamma_{kt}=\cos\gamma_{tk}$, this expression simplifies to
\begin{eqnarray}
\nonumber S_l^2&=&(4\pi)^2\left[\sum_{k=t}q_k^2\,g_l^2(\lambda_k)\right.\\
&&+\left.2\sum_{k>t}q_k\,q_t\,g_l(\lambda_k)\,g_l(\lambda_t)\,P_l(\cos\gamma_{kt})\right].
\end{eqnarray}
Normalizing this expression with $|S_0|=4\pi\sum_k|q_k|$, we obtain Eq.~\ref{eq:Sl}. Knowing the multipole magnitudes, we can also immediately write down the total power
\begin{equation}
\label{eq:pow}
P=\int_\Omega|\sigma(\Omega)|^2\mathrm{d}\Omega=\sum_l\frac{4\pi}{2l+1}\sum_m|\sigma_{lm}|^2=\sum_l S_l^2.
\end{equation}

\begin{comment}
As the form of the multipole magnitudes, Eq.~\eqref{eq:mag}, is similar to the form of the bond order parameters introduced by Steinhardt and Nelson~\cite{Steinhardt1983}, we can extend this analogy to form third-order invariants from the multipole coefficients,
\begin{equation}
\label{eq:Wl}
W_l=\frac{\sum_{m_1+m_2+m_3=0}\tj{l}{l}{l}{m_1}{m_2}{m_3}\sigma_{lm_1}\sigma_{lm_2}\sigma_{lm_3}}{\left(\sum_m|\sigma_{lm}^2|\right)^{3/2}},
\end{equation}
where the expression in the parentheses is the Wigner 3-j symbol.
%The ratio $S_l/W_l$ should also be relevant.
\end{comment}

\section{\label{sec:limits}Limiting cases}

In the limit where $\lambda\to0$, the asymptotic behavior of $g_l(\lambda)$ is given by
\begin{equation}
\lim_{\lambda\to0}g_l(\lambda)\propto\lambda^l+\mathcal{O}(\lambda^{l+2}),
\end{equation}
and specifically, $\lim_{\lambda\to0}g_0(\lambda)=1$. In this limit, the term with $l=0$ becomes dominant, and thus the only non-zero multipole coefficient $\sigma_{lm}$ is that with $l=m=0$. There, $\lim_{\lambda\to0}\sigma_{00}=\sqrt{4\pi}\,Q$, with $Q=\sum_kq_k$ the total charge on the sphere. Inserting this into the expression for the multipole expansion of the surface charge density, Eq.~\eqref{eq:multipole}, we get indeed that
\begin{equation}
\lim_{\lambda\to0}\sigma(\Omega)=\frac{Qe_0}{4\pi R^2},
\end{equation}
a uniform distribution on a sphere.

In the other limit where $\lambda\to\infty$, the function $g_l(\lambda)$ always tends to $1$, independent of $l$:
\begin{equation}
\lim_{\lambda\to\infty}g(l,\lambda)=1\quad\forall l.
\end{equation}
The multipole coefficients $\sigma_{lm}$ simplify to $\lim_{\lambda\to\infty}\sigma_{lm}=4\pi\sum_kq_k\,Y_{lm}^*(\Omega_k)$, yielding
\begin{eqnarray}
\nonumber\lim_{\lambda\to\infty}\sigma(\Omega)&=&\frac{e_0}{R^2}\sum_k q_k\sum_{l,m}Y_{lm}^*(\Omega_k)Y_{lm}(\Omega)\\
&=&\frac{e_0}{R^2}\sum_k q_k\,\delta(\Omega-\Omega_k),
\end{eqnarray}
which is indeed a surface charge distribution composed of Dirac $\delta$ functions centered at $\Omega_k$. Here we also see why, when using Dirac $\delta$ functions for the description of point charges, in principle an infinite sum over $\ell$ is needed to represent the distribution correctly.

As for the multipole magnitudes, the only non-zero moment in the limit of $\lambda\to0$ is of course $S_0$, with all higher multipoles tending to zero, $\lim_{\lambda\to0}S_l/|S_0|=0$ for $l\geqslant1$. On the other hand, the multipole magnitudes in the limit of $\lambda\to\infty$ are determined purely by their geometrical factor
\begin{equation}
\lim_{\lambda\to\infty}\frac{S_l}{|S_0|}=\frac{S_l^\infty}{|S_0|}=\sqrt{\frac{1}{N}+\frac{2}{N^2}\sum_{k>t}P_l(\cos\gamma_{kt})},
\end{equation}
which is dependent on the spherical distances between the charges on the sphere, $\cos\gamma_{kt}$.

\section{\label{sec:extra}Additional figures}

Here, we show several additional figures complementing the results presented in the main text.

\begin{figure}[!htp]
\includegraphics[width=\columnwidth]{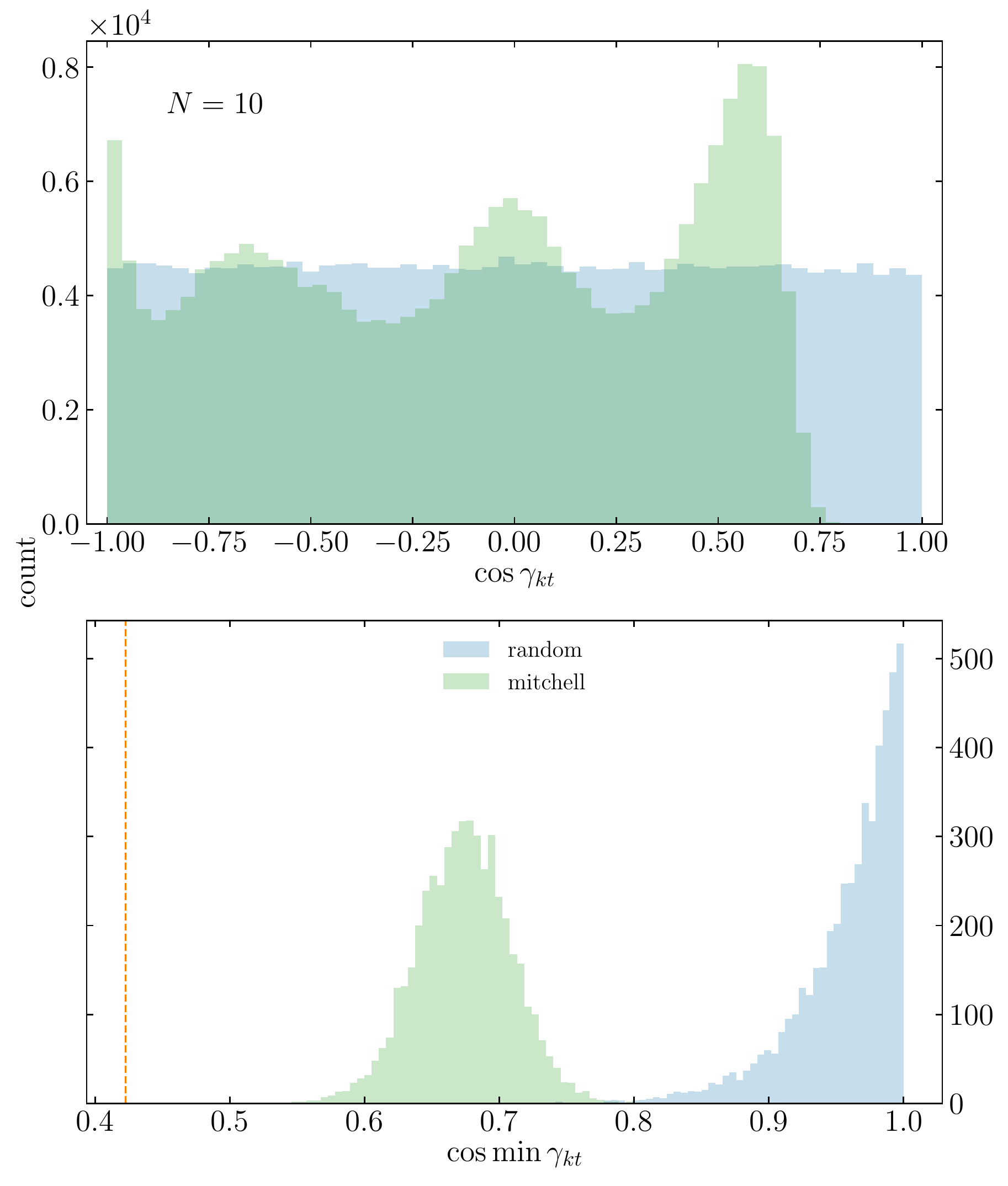}
\caption{Histogram of distances between charges, given by $\cos\gamma_{kt}$, and of the closest distance between two charges, $\cos\min\gamma_{kt}$ with $k\neq t$. The histograms were obtained from $5000$ different random and Mitchell configurations of $N=10$ identical charges. The dashed vertical line shows the minimum distance of the corresponding Thomson configuration.
\label{fig:E1}}
\end{figure}

\clearpage
\newpage

\setcounter{figure}{10}
\begin{widetext}
\begin{figure*}[]
\includegraphics[width=2\columnwidth]{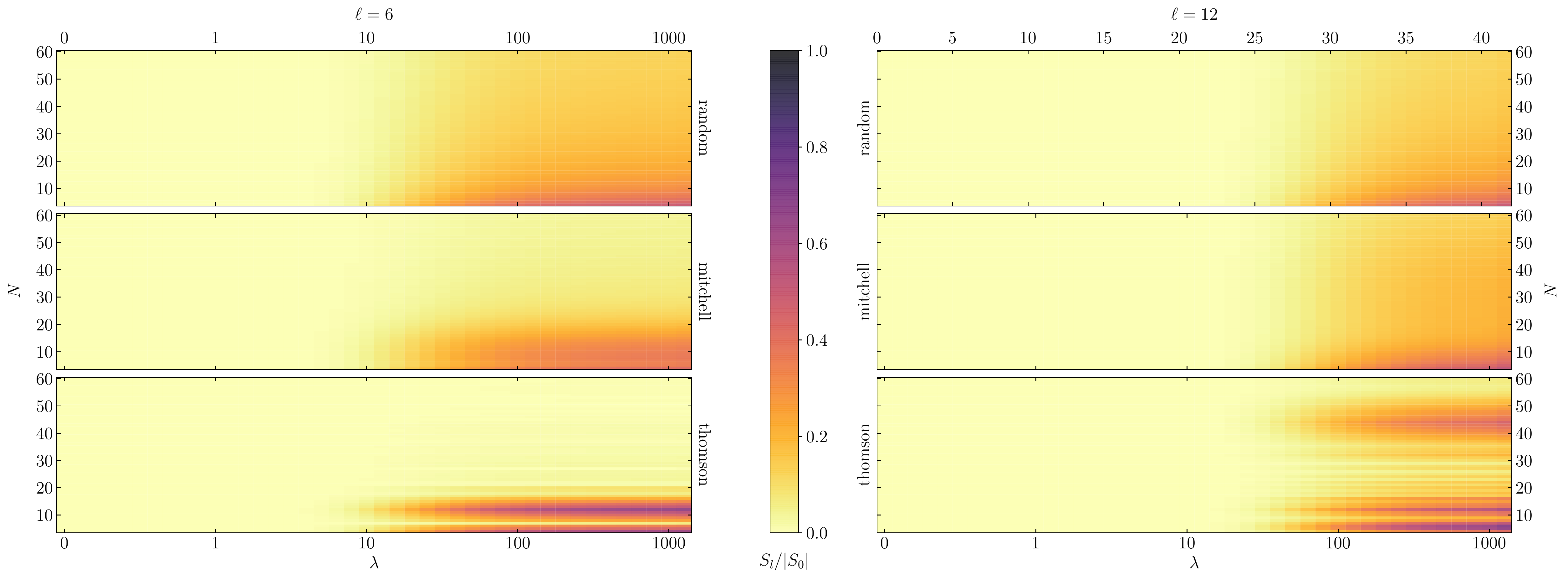}
\caption{Heatmaps of normalized multipole magnitudes $S_l/|S_0|$ as a function of $\lambda$ and $N$ for two different values of $\ell=6$, $12$. In the case of random and Mitchell configurations, the heatmaps show the mean values obtained by averaging over $5000$ different configurations. The last columns of the heatmaps show the values of the multipole magnitudes in the limit $\lambda\to\infty$, $S_l^\infty/|S_0|$.
\label{fig:E4}}
\end{figure*}
\end{widetext}

\setcounter{figure}{8}
\begin{figure}[!htp]
\includegraphics[width=\columnwidth]{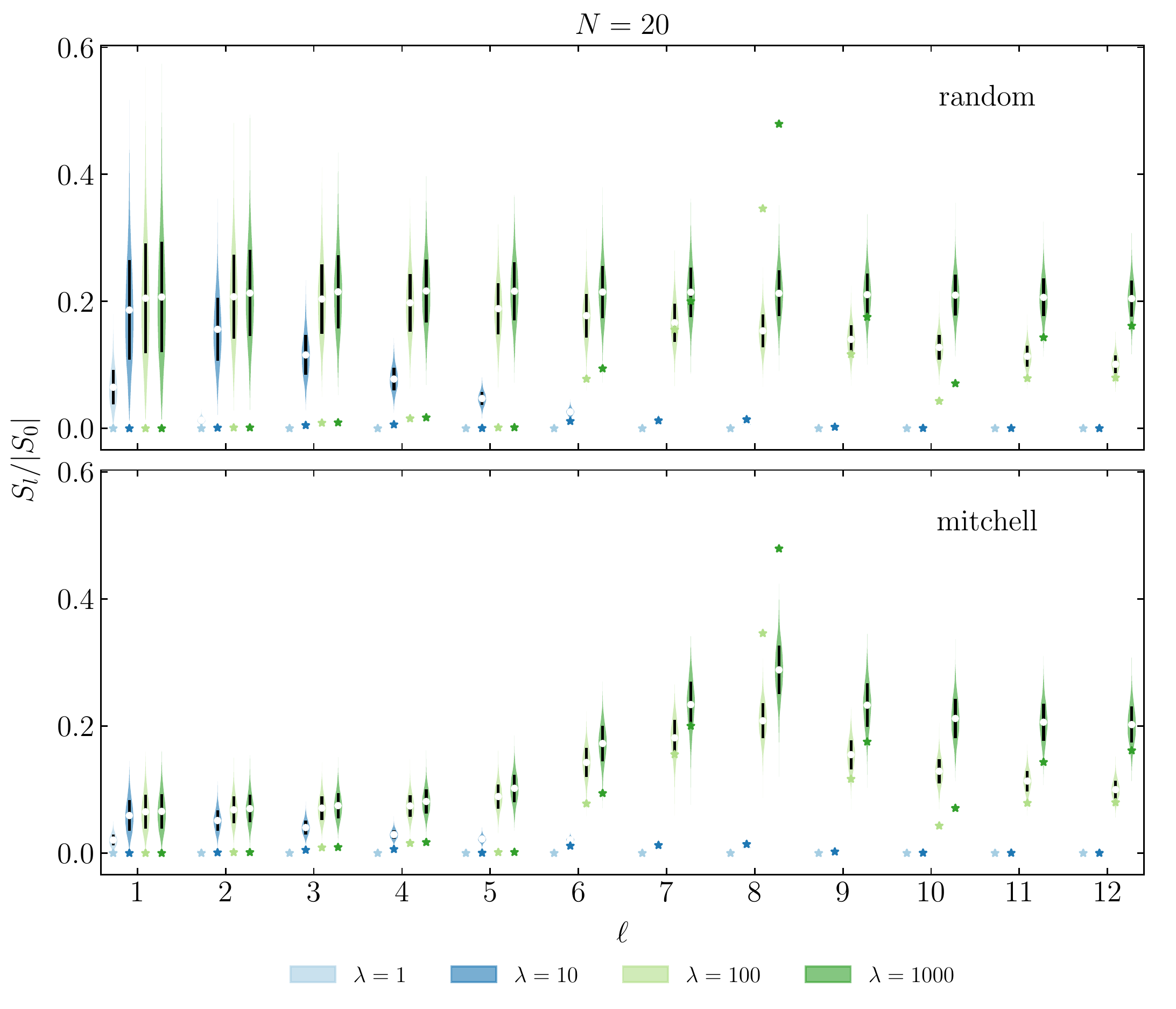}
\caption{Violin plot of the first $12$ multipole magnitudes for configurations of $N=20$ identical charges, generated either randomly or by using Mitchell's algorithm. Each entry in the violin plot shows a (mirrored) distribution of normalized magnitudes of $5000$ different configurations, with the central symbols denoting the mean and the bars denoting the corresponding standard deviation. Star symbols show the multipole magnitudes of the corresponding Thomson configuration. The plot is shown for four different values of the concentration parameter $\lambda$.
\label{fig:E2}}
\end{figure}

\begin{figure}[!htp]
\includegraphics[width=\columnwidth]{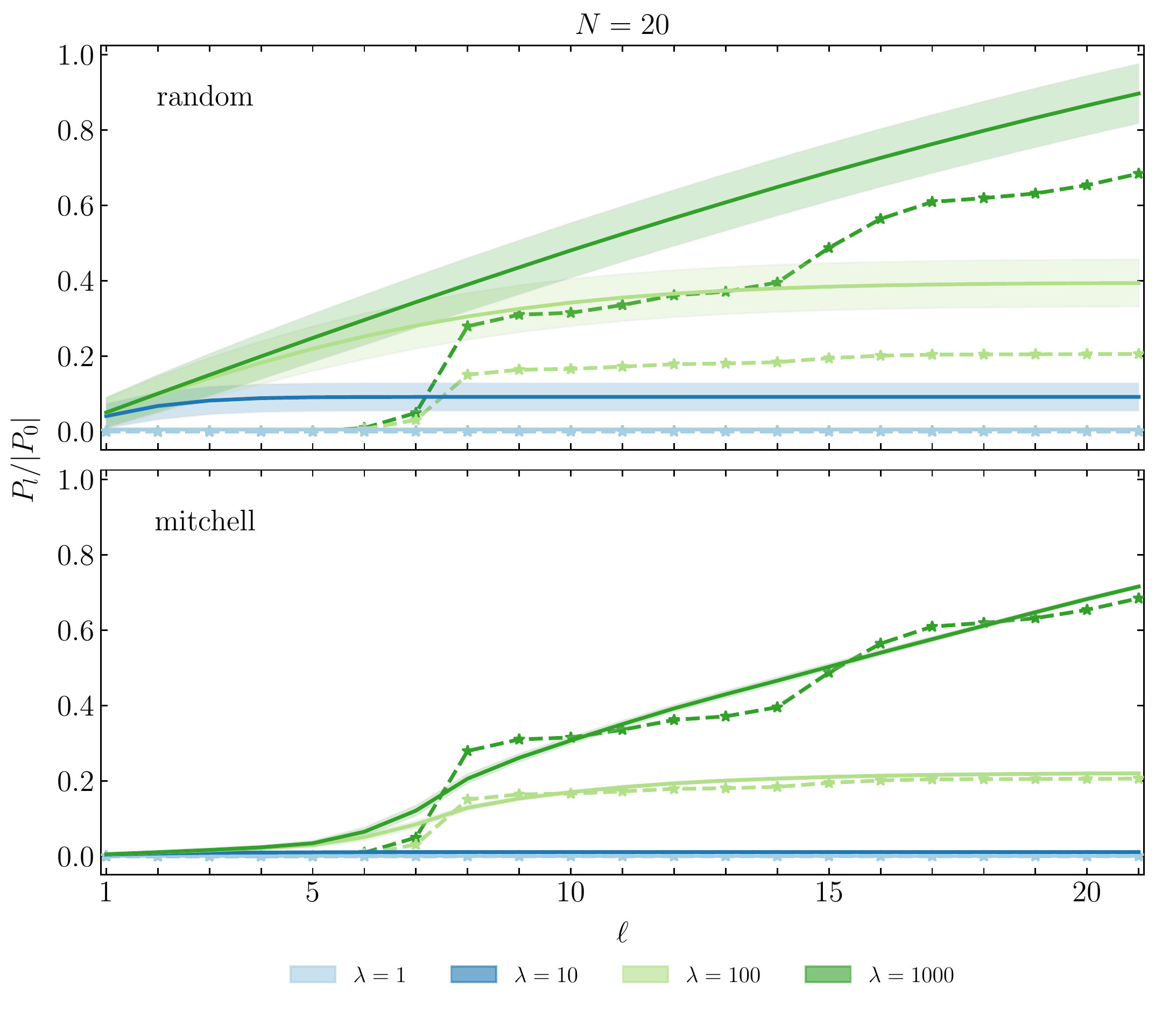}
\caption{Normalized total power $P_l/|P_0|$ of random and Mitchell configurations of $N=20$ identical charges, obtained by summing the squares of multipole magnitudes up to  order $\ell$. Full lines show the mean values, obtained by averaging over $5000$ different configurations, while the shaded regions denote the corresponding standard deviations. The latter are negligible for Mitchell configurations. Dashed lines and star symbols show the total power for the corresponding Thomson configuration. The plot is shown for four different values of the concentration parameter $\lambda$.
\label{fig:E3}}
\end{figure}

\clearpage

%\bibliographystyle{utphys}
%\bibliography{references}

\begin{thebibliography}{10}

\bibitem{Holm2012}
C.~Holm, P.~K{\'e}kicheff, and R.~Podgornik, eds., {\em Electrostatic Effects
  in Soft Matter and Biophysics}, vol.~46 of {\em NATO Science Series II --
  Mathematics, Physics and Chemistry}.
\newblock Springer, 2012.

\bibitem{Bianchi2017}
E.~Bianchi, B.~Capone, I.~Coluzza, L.~Rovigatti, and P.~D.~J. van Oostrum,
  {\em Limiting the valence: advancements and new perspectives on patchy
  colloids{,} soft functionalized nanoparticles and biomolecules}, Phys. Chem.
  Chem. Phys. {\bfseries 19}, 19847--19868 (2017).

\bibitem{Bianchi2017b}
E.~Bianchi, P.~D. van Oostrum, C.~N. Likos, and G.~Kahl,  {\em Inverse patchy
  colloids: Synthesis, modeling and self-organization}, Curr. Op. Colloid
  Interface Sci. {\bfseries 30}, 8--15 (2017).

\bibitem{Siber2012}
A.~\v{S}iber, A.~{Lo\v{s}dorfer Bo\v{z}i\v{c}}, and R.~Podgornik,  {\em
  Energies and pressures in viruses: contribution of nonspecific electrostatic
  interactions}, Phys. Chem. Chem. Phys. {\bfseries 14}, 3746--3765 (2012).

\bibitem{Bai2016}
Y.~Bai, Q.~Luo, and J.~Liu,  {\em Protein self-assembly via supramolecular
  strategies}, Chem. Soc. Rev. {\bfseries 45}, 2756--2767 (2016).

\bibitem{Bianchi2014}
E.~Bianchi, C.~N. Likos, and G.~Kahl,  {\em Tunable assembly of heterogeneously
  charged colloids}, Nano Lett. {\bfseries 14}, 3412 (2014).

\bibitem{Barisik2014}
M.~Barisik, S.~Atalay, A.~Beskok, and S.~Qian,  {\em Size dependent surface
  charge properties of silica nanoparticles}, J. Phys. Chem. C {\bfseries 118},
  1836--1842 (2014).

\bibitem{Kusters2015}
R.~Kusters, H.-K. Lin, R.~Zandi, I.~Tsvetkova, B.~Dragnea, and P.~van~der
  Schoot,  {\em Role of charge regulation and size polydispersity in
  nanoparticle encapsulation by viral coat proteins}, J. Phys. Chem. B
  {\bfseries 119}, 1869--80 (2015).

\bibitem{Sabapathy2017}
M.~Sabapathy, R.~A.~M. K, and E.~Mani,  {\em Self-assembly of inverse patchy
  colloids with tunable patch coverage}, Phys. Chem. Chem. Phys. {\bfseries
  19}, 13122--13132 (2017).

\bibitem{ALB2017a}
A.~{Lo\v{s}dorfer Bo\v{z}i\v{c}} and R.~Podgornik,  {\em {pH} dependence of
  charge multipole moments in proteins}, Biophys. J. {\bfseries 113},
  1454--1465 (2017).

\bibitem{Krishnan2017}
M.~Krishnan,  {\em A simple model for electrical charge in globular
  macromolecules and linear polyelectrolytes in solution}, J. Chem. Phys.
  {\bfseries 146}, 205101 (2017).

\bibitem{Nap2014}
R.~J. Nap, A.~{Lo\v{s}dorfer Bo\v{z}i\v{c}}, I.~Szleifer, and R.~Podgornik,
  {\em The role of solution conditions in the bacteriophage {PP7} capsid charge
  regulation}, Biophys. J. {\bfseries 107}, 1970--1979 (2014).

\bibitem{Abrikosov2017}
A.~I. Abrikosov, B.~Stenqvist, and M.~Lund,  {\em Steering patchy particles
  using multivalent electrolytes}, Soft Matter  (2017).

\bibitem{Ni2012}
P.~Ni, Z.~Wang, X.~Ma, N.~C. Das, P.~Sokol, W.~Chiu, B.~Dragnea, M.~Hagan, and
  C.~C. Kao,  {\em An examination of the electrostatic interactions between the
  N-terminal tail of the brome mosaic virus coat protein and encapsidated
  RNAs}, J. Mol. Biol. {\bfseries 419}, 284--300 (2012).

\bibitem{Warshel2006}
A.~Warshel, P.~K. Sharma, M.~Kato, and W.~W. Parson,  {\em Modeling
  electrostatic effects in proteins}, Biochim. Biophys. Acta Proteins
  Proteomics {\bfseries 1764}, 1647--1676 (2006).

\bibitem{Adar2017}
R.~M. Adar, D.~Andelman, and H.~Diamant,  {\em Electrostatics of patchy
  surfaces}, Adv. Colloid Interface Sci. {\bfseries 247}, 198--207 (2017).

\bibitem{Grant2001}
M.~Grant,  {\em Nonuniform charge effects in protein- protein interactions}, J.
  Phys. Chem. B {\bfseries 105}, 2858--2863 (2001).

\bibitem{ALB2013a}
A.~{Lo\v{s}dorfer Bo\v{z}i\v{c}} and R.~Podgornik,  {\em Symmetry effects in
  electrostatic interactions between two arbitrarily charged shells in the
  {D}ebye-{H}\"{u}ckel approximation}, J. Chem. Phys. {\bfseries 138}, 074902
  (2013).

\bibitem{Li2017}
S.~Li, G.~Erdemci-Tandogan, J.~Wagner, P.~van~der Schoot, and R.~Zandi,  {\em
  Impact of a nonuniform charge distribution on virus assembly}, Phys. Rev. E
  {\bfseries 96}, 022401 (2017).

\bibitem{Li2015}
W.~Li, B.~A. Persson, M.~Morin, M.~A. Behrens, M.~Lund, and
  M.~Zackrisson~Oskolkova,  {\em Charge-induced patchy attractions between
  proteins}, J. Phys. Chem. B {\bfseries 119}, 503--508 (2015).

\bibitem{Vega2016}
J.~F. Vega, E.~Vicente-Alique, R.~Núñez-Ramírez, Y.~Wang, and
  J.~Martínez-Salazar,  {\em Evidences of Changes in Surface Electrostatic
  Charge Distribution during Stabilization of HPV16 Virus-Like Particles}, PLoS
  ONE {\bfseries 11}, 1--17 (2016).

\bibitem{Blanco2016}
M.~A. Blanco and V.~K. Shen,  {\em Effect of the surface charge distribution on
  the fluid phase behavior of charged colloids and proteins}, J. Chem. Phys.
  {\bfseries 145}, 155102 (2016).

\bibitem{Dempster2016}
J.~M. Dempster and M.~Olvera de~la Cruz,  {\em Aggregation of heterogeneously
  charged colloids}, ACS Nano {\bfseries 10}, 5909--5915 (2016).

\bibitem{Yigit2015}
C.~Yigit, J.~Heyda, and J.~Dzubiella,  {\em Charged patchy particle models in
  explicit salt: Ion distributions, electrostatic potentials, and effective
  interactions}, J. Chem. Phys. {\bfseries 143}, 064904 (2015).

\bibitem{Yigit2017}
C.~Yigit, M.~Kandu\v{c}, M.~Ballauff, and J.~Dzubiella,  {\em Interaction of
  Charged Patchy Protein Models with Like-Charged Polyelectrolyte Brushes},
  Langmuir {\bfseries 33}, 417--427 (2017).

\bibitem{Silbert2012}
G.~Silbert, D.~Ben-Yaakov, Y.~Dror, S.~Perkin, N.~Kampf, and J.~Klein,  {\em
  Long-Ranged Attraction between Disordered Heterogeneous Surfaces}, Phys. Rev.
  Lett. {\bfseries 109}, 168305 (2012).

\bibitem{Perkin2006}
S.~Perkin, N.~Kampf, and J.~Klein,  {\em Long-Range Attraction between
  Charge-Mosaic Surfaces across Water}, Phys. Rev. Lett. {\bfseries 96}, 038301
  (2006).

\bibitem{Meyer2005}
E.~E. Meyer, Q.~Lin, T.~Hassenkam, E.~Oroudjev, and J.~N. Israelachvili,  {\em
  Origin of the long-range attraction between surfactant-coated surfaces},
  Proc. Natl. Acad. Sci. USA {\bfseries 102}, 6839--6842 (2005).

\bibitem{Hoppe2013}
T.~Hoppe,  {\em A simplified representation of anisotropic charge distributions
  within proteins}, J. Chem. Phys. {\bfseries 138}, 174110 (2013).

\bibitem{Stipsitz2015}
M.~Stipsitz, G.~Kahl, and E.~Bianchi,  {\em Generalized inverse patchy colloid
  model}, J. Chem. Phys. {\bfseries 143}, 114905 (2015).

\bibitem{Felder2007}
C.~E. Felder, J.~Prilusky, I.~Silman, and J.~L. Sussman,  {\em A server and
  database for dipole moments of proteins}, Nucleic Acids Res. {\bfseries 35},
  W512--W521 (2007).

\bibitem{Nakamura1985}
H.~Nakamura and A.~Wada,  {\em Nature of the charge distribution in proteins.
  III. Electric multipole structures}, J. Phys. Soc. Jpn. {\bfseries 54},
  4047--4052 (1985).

\bibitem{Paulini2005}
R.~Paulini, K.~M{\"u}ller, and F.~Diederich,  {\em Orthogonal multipolar
  interactions in structural chemistry and biology}, Angew. Chem. Int. Ed.
  {\bfseries 44}, 1788--1805 (2005).

\bibitem{Parimal2014}
S.~Parimal, S.~M. Cramer, and S.~Garde,  {\em Application of a spherical
  harmonics expansion approach for calculating ligand density distributions
  around proteins}, J. Phys. Chem. B {\bfseries 118}, 13066--13076 (2014).

\bibitem{Kim2006}
J.~Y. Kim, S.~H. Ahn, S.~T. Kang, and B.~J. Yoon,  {\em Electrophoretic
  mobility equation for protein with molecular shape and charge multipole
  effects}, J. Colloid Interface Sci. {\bfseries 299}, 486--492 (2006).

\bibitem{Gramada2006}
A.~Gramada and P.~E. Bourne,  {\em Multipolar representation of protein
  structure}, BMC Bioinformatics {\bfseries 7}, 242 (2006).

\bibitem{Lorman2007}
V.~Lorman and S.~Rochal,  {\em Density-wave theory of the capsid structure of
  small icosahedral viruses}, Phys. Rev. Lett. {\bfseries 98}, 185502 (2007).

\bibitem{Lorman2008}
V.~Lorman and S.~Rochal,  {\em Landau theory of crystallization and the capsid
  structures of small icosahedral viruses}, Phys. Rev. B {\bfseries 77}, 224109
  (2008).

\bibitem{ALB2011}
A.~{Lo\v{s}dorfer Bo\v{z}i\v{c}}, A.~\v{S}iber, and R.~Podgornik,  {\em
  Electrostatic self-energy of a partially formed spherical shell in salt
  solution: Application to stability of tethered and fluid shells as models for
  viruses and vesicles}, Phys. Rev. E {\bfseries 83}, 041916 (2011).

\bibitem{Mardia2009}
K.~V. Mardia and P.~E. Jupp, {\em Directional statistics}, vol.~494.
\newblock John Wiley \& Sons, 2009.

\bibitem{Abramowitz}
M.~Abramowitz and I.~A. Stegun, {\em Handbook of mathematical functions},
  vol.~55.
\newblock Dover Publications, 1964.

\bibitem{Mitchell1991}
D.~P. Mitchell,  {\em Spectrally Optimal Sampling for Distribution Ray
  Tracing}, SIGGRAPH Comput. Graph. {\bfseries 25}, 157--164 (1991).

\bibitem{Wales2006}
D.~J. Wales and S.~Ulker,  {\em Structure and dynamics of spherical crystals
  characterized for the Thomson problem}, Phys. Rev. B {\bfseries 74}, 212101
  (2006).

\bibitem{Snyder}
J.~P. Snyder, {\em Map projections -- A working manual}, vol.~1395.
\newblock US Government Printing Office, Washington, DC, 1987.

\bibitem{Gelessus1995}
A.~Gelessus, W.~Thiel, and W.~Weber,  {\em Multipoles and symmetry}, J. Chem.
  Ed. {\bfseries 72}, 505 (1995).

\bibitem{ALB2013b}
A.~{Lo\v{s}dorfer Bo\v{z}i\v{c}}, A.~\v{S}iber, and R.~Podgornik,  {\em
  Statistical analysis of sizes and shapes of virus capsids and their resulting
  elastic properties}, J. Biol. Phys. {\bfseries 39}, 215--228 (2013).

\bibitem{Stone1981}
A.~Stone,  {\em Distributed multipole analysis, or how to describe a molecular
  charge distribution}, Chem. Phys. Lett. {\bfseries 83}, 233--239 (1981).

\bibitem{Larsson1985}
S.~Larsson and M.~Braga,  {\em Atomic charges based on spherical harmonics
  expansion at the atomic centers}, Theor. Chim. Acta {\bfseries 68}, 291--300
  (1985).

\bibitem{Dharmavaram2017}
S.~Dharmavaram, F.~Xie, W.~Klug, J.~Rudnick, and R.~Bruinsma,  {\em
  Orientational phase transitions and the assembly of viral capsids}, Phys.
  Rev. E {\bfseries 95}, 062402 (2017).

\bibitem{Kent1982}
J.~T. Kent,  {\em The Fisher-Bingham distribution on the sphere}, J. R. Stat.
  Soc. Series B Stat. Methodol. {\bfseries 44}, 71--80 (1982).

\bibitem{Arfken}
G.~B. Arfken and H.~J. Weber, {\em Mathematical methods for physicists}.
\newblock Academic Press, San Diego, CA, 4th~ed., 1995.

\bibitem{Mathematica}
{Wolfram Research Inc.}, {\em Mathematica 8.0}, 2010.

\end{thebibliography}
\providecommand{\href}[2]{#2}\begingroup\raggedright\endgroup

\end{document}